\documentclass[11pt,usletter]{article}
\usepackage{jheppub}
\usepackage{hyperref}
\usepackage{amsmath}
\usepackage{amsfonts}
\usepackage{graphicx}
\usepackage{verbatim}
\usepackage{indentfirst}
\usepackage{eufrak}
\usepackage{xcolor}
\usepackage{mathtools}
\usepackage{subfigure}
\usepackage{tikz}
\tikzset{every picture/.style={line width=0.75pt}} 

\renewcommand{\[}{\begin{equation}}
\renewcommand{\]}{\end{equation}}

\def\XXint#1#2#3{{\setbox0=\hbox{$#1{#2#3}{\int}$}
     \vcenter{\hbox{$#2#3$}}\kern-.5\wd0}}

\def\nfrac#1#2{\genfrac{}{}{0pt}{}{#1}{#2}}

\usepackage[OT2,T1]{fontenc}
\DeclareSymbolFont{cyrletters}{OT2}{wncyr}{m}{n}

\title{On the analyticity of the lightest particle mass\\
of Ising field theory in a magnetic field}
\author[a]{Hao-Lan Xu}

\affiliation[a]{C.N. Yang Institute for Theoretical Physics, State University of New York, Stony Brook, NY 11794-3840, USA}

\emailAdd{hao-lan.xu@stonybrook.edu}

\begin{document}
\bibliographystyle{unsrt}

\begin{flushright}
YITP-SB-2024-10\\
\end{flushright}
\abstract{
We study the scaling functions associated with the lightest particle mass $M_1$ in 2d Ising field theory in external magnetic field. The scaling functions depend on the scaling parameter $\xi = h/|m|^{\frac{15}{8}}$, or related parameter $\eta = m / h^{\frac{8}{15}}$. Analytic properties of $M_1$ in the high-T domain were discussed in \cite{Xu:2022mmw}. In this work, we study analyticity of $M_1$ in the low-T domain. Important feature of this analytic structure is represented by the Fisher-Langer's branch cut. The discontinuity across this branch cut determines the behavior of $M_1$ at all complex $\xi$ via associated low-T dispersion relation. Also, we put forward the "extended analyticity" conjecture for $M_1$ in the complex $\eta$-plane, similar to the analyticity of the free energy density previously proposed in \cite{fonseca2003ising}. The extended analyticity implies the "extended dispersion relation", which we verify against the numerics from the Truncated Free Fermion Approach (TFFSA), giving strong support to the conjecture.


}
\maketitle
\flushbottom


\vskip 1cm
\section{Introduction}\label{Section1}
The scaling behaviour of Ising model near its ferromagnetic critical point describes one of the most important universality classes of two dimensional criticality, and the related field theory interpretation, known as 2d Ising Field Theory (IFT), is an interesting example of 2d quantum field theory with rich particle spectrum and S-matrices. In this report, as a continuation of previous works \cite{fonseca2003ising}\cite{zamolodchikov2013ising}\cite{Xu:2022mmw}\cite{Xu:2023nke}, we focus on the mass of lightest excitation, denoted as $M_1$, and discuss its analytic properties as the function of the scaling parameter.

\subsection*{2d Ising field Theory}
At the ferromagnetic critical point $(T,H) = (T_c, 0)$, say at the Curie temperature with no external magnetic field, the scaling limit of Ising model is described by Ising Conformal Field Theory (ICFT), which in two dimension is represented by the minimal CFT $\mathcal M_{3,4},$ with Virasoro central charge $c_{3,4} = \frac{1}{2}$. From Wilsonian point of view, the Ising ferromagnetic critical point is a fix point of Renormalization Group (RG) flows. Near the Ising critical point, the scaling limit of 2d Ising model is described by 2d Ising field theory, which alternatively can be understood in terms of the RG flow initiated from the Ising fixed point, generated by perturbing with two relevant operators. The description is given by the formal action:
\begin{equation}
\mathcal A_{\text{IFT}} = \mathcal A_{\text{FF}} + h \int \sigma(x)\, d^2 x  = \mathcal A^{*}_{3,4} + \frac{m}{2 \pi }\int \varepsilon(x)\, d^2 x + h \int \sigma(x)\, d^2 x \, ,\label{IFTaction}
\end{equation}
where $\mathcal A^{*}_{3,4}$ is the action of minimal model $\mathcal M_{3,4}$, $\sigma(x)$ and $\varepsilon(x)$ are local spin density and energy density operators respectively. The couplings $m$ and $h$ are related to the infinitesimal deviation from the critical point, say:
\[
m \propto \frac{T_c - T}{T_c} \,, \quad \text{and} \quad h \propto H \,, \label{mThHrelation}
\]
with the normalization coefficients are fixed by requiring the short-distance behaviours\footnote{The proportionality coefficient in \eqref{mThHrelation} depends on specific microscopic model, for details, see e.g. \cite{BazhanovYL}.}:
\begin{equation}
 |x|^{\frac{1}{4}} \langle \sigma(x) \sigma(0) \rangle \to 1 \,, \quad   |x|^2 \langle \varepsilon(x) \varepsilon(0) \rangle \to 1
\end{equation}
at $x \to 0$. Note that the two nontrivial local relevant scalar operators $\sigma(x)$ and $\varepsilon(x)$ have the scaling dimensions $(\Delta_\sigma, \bar{\Delta}_\sigma) = (\frac{1}{16},\frac{1}{16})$ and $(\Delta_\varepsilon, \bar{\Delta}_\varepsilon) = (\frac{1}{2},\frac{1}{2})$ respectively, hence the corresponding relevant couplings have mass dimensions $[h] = \frac{15}{8}$ and $[m] = 1$. The dimensionless combination
\begin{equation}
\xi = \frac{h}{|m|^{\frac{15}{8}}} \quad \Big( \quad \text{or} \quad \eta = \frac{m}{h^{\frac{8}{15}}} \quad \Big) \,, \label{scalingparameterdef}
\end{equation}
\footnote{Depending on the sign of $m$, $\eta$ is related to $\xi$ by $\eta = \xi^{-8/15}$ in the low-T regime $m>0$, or $\eta =-\xi^{-8/15}$ in the high-T regime $m<0$.}labels the renormalization group trajectories stemming from the Ising CFT fixed point. As is conventional, we will call $\xi$ (or $\eta$) the scaling parameter.

The scaling behaviour of the IFT is described by dimensionless functions known as scaling functions, which only depend on the scaling parameter $\xi$ or $\eta$. In this work, we study the mass of the lightest particle, denoted as $M_1=M_1(m,h)$. As the energy of lightest excitation, $M_1$ carries some of the most important information of the theory. In particular, $M_1$ can be identified with the inverse of correlation length, as $R_c = M_1^{-1}$, and criticality occurs when $M_1 \to 0$. The scaling functions of $M_1$ are defined as:
\[
\hat{M}_1 = \hat{M}_1(\xi) = {M_1}/{|m|} \,,
\quad \text{and} \quad
\mathcal M_1 = \mathcal M_1(\eta) =  {M_1}/{|h|^{\frac{8}{15}}} \,.
\]
Both functions $\hat M_1(\xi)$ and $\mathcal M_1(\eta)$ can be analytically continued to complex values of the scaling parameters $\xi$ or $\eta$. It is the analytic properties of these scaling functions as the functions of complex $\xi$ and $\eta$ which are studied in this work. Part of the analyticity of $\hat{M}_1(\xi)$ was discussed in \cite{Xu:2022mmw} in the $m < 0$ high-T regime, while in this work, we will discuss the analyticity of $\hat M_1(\xi)$ in the $m > 0$ low-T regime, and the analyticity of $\mathcal M_1(\eta)$ on the complex $\eta$-plane, which unifies the high-T and low-T analyticities of $\hat M_1(\xi)$. We expect the analyticities of $\hat M_1(\xi)$ and $\mathcal M_1(\eta)$ to be similar to the scaling functions associated with the vacuum energy density $F(m,h)$, which were analyzed in \cite{fonseca2003ising}.


\subsection*{High-T analyticity of $M_1$}
\begin{figure}[!h]
\centering
\includegraphics[width=0.575\textwidth]{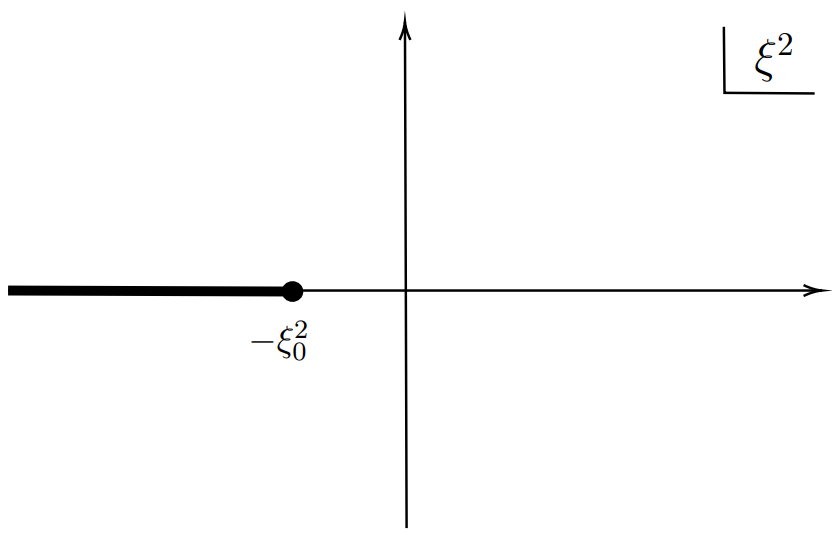}
\caption{Phase diagram of $\hat{M}_1(u)$ on the complex $u$-plane, showing high-T standard analyticity. The Yang-Lee singularity is located at $u = - \xi_0^2$, and we choose the Yang-Lee branch cut extends from $u = -\xi_0^2$ to $ u = -\infty$. The Yang-Lee branch cut represents a line of first order phase transition with complex state functions.
}
\label{PhaseDiagramXi}
\end{figure}

In this subsection, we briefly review the high-T analyticity of $\hat M_1(\xi)$, as discussed in \cite{Xu:2022mmw}. In the high-T regime $T > T_c$ (equivalently  $m < 0$), IFT enjoys the $\mathbb Z_2$ symmetry $h \to -h,\ \sigma(x) \to - \sigma(x)$. As the result, all scaling functions should be even functions of $\xi$. It is conventional to denote instead the scaling function as $\hat M_1 = \hat M_1(u)$, where $u := \xi^2$. From $u>0$, $\hat{M}_1(u)$ can be analytically continued to generic complex values of $u$. $\hat{M}_1(u)$ is expected to be analytic everywhere on the complex $u$-plane, except at the Yang-Lee (YL) branching point and the associated YL branch cut. The Yang-Lee branching point, also known as Yang-Lee edge singularity, is located at $\xi^2 = - \xi_0^2 \approx -0.03583 \dots$ \cite{fonseca2003ising}\cite{Xu:2022mmw}\cite{BazhanovYL}. As in \cite{fonseca2003ising}\cite{Xu:2022mmw}, we chose the branch cut to extend along the real $u$-axis, from $u = -\infty$ to $u = -\xi_0^2$. This defines the principle branch of $\hat M_1(u)$. The Yang-Lee branch cut represents a line of first order phase transition with complex state functions. The analyticity structure is as shown in Fig.\ref{PhaseDiagramXi}.

In fact, in high-T many other scaling functions have similar analyticities on the complex $u$-plane, as in Fig.\ref{PhaseDiagramXi}. For example, the scaling functions of free energy density and effective "$\varphi^3$ coupling" were discussed in \cite{fonseca2003ising} and \cite{Xu:2023nke}. This is known as "high-T standard analyticity", which is resulting from the Yang-Lee theorem of Ising ferromagnetic partition function \cite{yang1952statistical}\cite{lee1952statistical}. On the lattice, define the fugacity $\mu = e^{-2 H / k_B T}$, the Yang-Lee theorem states that zeros of lattice Ising partition function are distributed along the unit circle of the $\mu$-plane, known as the Yang-Lee circle. At $T > T_c$ there is a segment $| \text{Arg}\, \mu | \le \text{Arg}\, \mu_0$ free of any zeros, and the zeros are located on the circle with $| \text{Arg}\, \mu | > \text{Arg}\, \mu_0$. In the scaling limit with $T - T_c \to 0_+$ and $H \to 0$, the zeros condense into the Yang-Lee branch cut on the complex $u$-plane, which represents a line of first order phase transition with complex state functions\footnote{With the choice of Yang-Lee branch cut as in Fig.\ref{PhaseDiagramXi}, the scaling functions come as complex conjugate pairs on both edges of branch cut. For example, we denote $\hat{M}_1(u \pm i0) = \hat{M}_1^{(\pm)}(u)$ when $u < -\xi^2_0$, and $(\hat{M}_1^{(+)})^* = \hat{M}_1^{(-)}$.}. The edge of the condensing zeros becomes the Yang-Lee edge singularity, whose location is denoted as $u_0 = -\xi^2_0$. The Yang-Lee edge singularity represents a continuous phase transition, thus is also known as Yang-Lee critical point. The corresponding conformal field theory is the minimal model $\mathcal M_{2,5}$ \cite{cardy1985conformal}, usually referred to as the Yang-Lee conformal field theory (YLCFT). This CFT is non-unitary, with central charge $c_{2,5} = -\frac{22}{5}$.

Following the standard analyticity as in Fig.\ref{PhaseDiagramXi}, $\hat{M}_1(u)$ at any complex $u$ can be expressed through the discontinuities across the Yang-Lee branch cut. On the Yang-Lee branch cut, $\text{Disc}\, {\hat M_1} (u) = 2i\,\Im m\, {\hat M_1}(u)$, and we have
\[
\hat M_1( u ) = 1 + u \int_{\xi_0^2}^{\infty} \frac{dx}{\pi} \frac{\Im m \, \hat M_1(-x + i0)}{x (x + u)} \,, \label{M1highTDisp}
\]
which is known as the high-T mass dispersion relation \cite{Xu:2022mmw}. Similar dispersion relations also can be formulated for some other scaling functions, see e.g. \cite{mccoy2013two}, or \cite{fonseca2003ising}\cite{Xu:2023nke}. Additionally, the standard analyticity states that no other singularity exist along the YL branch cut, thus $\Im m \, \hat M_1(u + i0)$ is a smooth function on the interval $ -\infty< u \le -\xi^2_0$. The behaviour of discontinuity near both ends ($u = -\xi^2_0$ and $u = -\infty$) are controlled by the expansions in $(u + \xi_0^2)$ or $\eta = u^{-\frac{4}{15}}$.

When $u$ is close to $ -u_0 : = -\xi^2_0$, the Yang-Lee critical point serves as an infrared fixed point of renormalization group flow. The criticality are controlled by YLCFT, and the theory at infrared can be described by the effective action:
\[
\mathcal A_{\text{eff}} = \mathcal A^{*}_{2,5} + \lambda(u) \int \phi(x) d^2 x + \sum_i g_i(u) \int \mathcal O_i(x) d^2 x \,, \label{YLeffAction}
\]
where $\mathcal A^{*}_{2,5}$ is the action of YLCFT. $\phi(x)$ is the only nontrivial relevant scalar operator of minimal model $\mathcal M_{2,5}$, with normalization\footnote{In literature there are different choices of normalization, trading positivity norm of Hilbert space to real structure constant. By choosing the following $\phi(x)$, the structure constant $\mathbb C^{\phi}_{\phi\phi}$ is real. See e.g. \cite{fisher1978yang}\cite{Zamolodchikov:1990bk}\cite{Xu:2022mmw}.}:
\[
\langle \phi(x) \phi(0) \rangle \to - |x|^{4/5} \,,
\]
at $x \to 0$, and conformal dimensions $(\Delta_\phi,\bar{\Delta}_\phi) = (-\frac{1}{5} , -\frac{1}{5})$. $\lambda(u)$ is the associated relevant coupling with the mass dimension $[\lambda] = \frac{12}{5}$ \footnote{When $u > -\xi^2_0$, $\lambda$ is positive, and the effective description has a unique vacuum.}. $\mathcal O_i(x)$ denote irrelevant scalar operators, which are scalar descendent operators in minimal model $\mathcal M_{2,5}$, and the corresponding couplings $g_i$'s have negative mass dimensions.

The effective action \eqref{YLeffAction} is infrared asymptotically integrable, i.e. it is integrable if one drops all the irrelevant operators of \eqref{YLeffAction}. The theory with action:
\[
\mathcal A_{\text{YLQFT}} = \mathcal A^{*}_{2,5} + \lambda \int \phi(x) d^2 x \, \label{YLQFTAction}
\]
is known as Yang-Lee QFT, as the deformation of YLCFT with its only relevant nontrivial scalar operator $\phi(x)$. The Yang-Lee QFT is gapped and integrable, with the solvable elastic S-matrix \cite{Cardy:1989fw}. But the effective description \eqref{YLeffAction} is not integrable at any finite energy scale, since with nonvanishing $g_i$ the generic irrelevant perturbations break the integrability\footnote{The exceptions which do not break integrability are the $T\bar T$ or $T\bar T$-like operators \cite{smirnov2017space}, but any descendant of $\phi(x)$ would break integrability, see \cite{Xu:2022mmw} for the analysis regarding $L_{-4} \bar L_{-4} \phi(x)$.}.

Denote the conformal dimension of the scalar $\mathcal O_i(x)$ as $(\Delta_i , \Delta_i)$, and the mass dimension of its coupling as $[g_i] = 2 - 2 \Delta_i$. The couplings admit regular expansion in powers of $\Delta u = u + u_0$:
\begin{gather}
\lambda(u) = \lambda_1 \,\Delta u + \lambda_2 \,\Delta u^2 + \cdots \,,\\ \label{LambdaExpansion}
g_i(u) =g_i^{(0)} + g_i^{(1)} \Delta u + g_i^{(2)} \Delta u^2 + \cdots \,.
\end{gather}
By dimensional analysis, the leading critical behaviour of $\hat M_1(u)$ is $\hat M_1 \sim \lambda^{5/12} \sim (u + u_0)^{5/12}$, which is determined by the relevant operator $\phi(x)$. At $u > -u_0$, the leading critical amplitude is given by $\hat M_1 / (u + u_0)^{\frac{5}{12}}  \to b_0 $ when approaching the Yang-Lee point. Beyond leading order, the form of subleading critical behaviors are also given by dimensional analysis:
\[
\frac{\hat M_1 (u)}{ b_0 (u + u_0)^{\frac{5}{12}}} = 1 + \sum_i C_{i} g_i(u)  \hat M_{\text{YL}}(u)^{2\Delta_i - 2} + \sum_{ij} C_{ij} g_i(u) g_j(u) \hat M_{\text{YL}}(u)^{2\Delta_i + 2\Delta_j - 4} + \cdots \,, \nonumber
\]
where $\hat M_{\text{YL}}(u) =C_{\text{YL}} \lambda(u)^{5/12} $ is the mass of Yang-Lee particle, where $C_{\text{YL}} = 2.6429 \dots$\cite{Zamolodchikov:1995xk} is the coefficient of Yang-Lee mass-coupling relation. $C_i$ and $C_{ij}$ are dimensionless numbers given by corresponding matrix elements\footnote{The couplings $C_{ij}$ and higher ones can not be determined via straightforward perturbation theory due to non-renormalizability, due to the presence of an infinite number of ambiguous counter terms. However, the general form of the expansion is fully determined by the dimensional analysis.}. For example, the lowest irrelevant scalar operator is the famous $T \bar T$ operator, with conformal dimensions $(2,2)$. As a result, the singular expansion of $\hat M_1$ reads:
\[
\hat M_1(u) = (u + u_0)^{\frac{5}{12}} \Big[ b_0 + b_1 (u + u_0 ) + c_0 (u + u_0)^{\frac{5}{6}} + \cdots \Big] \,. \label{M1SingularExpansion1}
\]
Here $b_1$ is related to $\lambda_2$ of \eqref{LambdaExpansion}, and $c_0$ is related to the leading $T\bar T$ coupling $g^{(0)}_{T \bar T}$. The critical amplitudes of \eqref{M1SingularExpansion1} were measured in \cite{Xu:2022mmw}. Below the Yang-Lee point with $u < - u_0$, analytic continuation of \eqref{M1SingularExpansion1} gives the behaviour of discontinuity near the Yang-Lee point, which reads (with $u + u_0 <0$):
\[
\Im m\, \hat M_1(u + i0) = (-u - u_0)^{\frac{5}{12}} \Big[ b_0 \sin\big( \frac{5\pi}{12} \big)+ b_1 ( - u - u_0 ) \sin\big( \frac{17\pi}{12}\big) + c_0 (- u - u_0)^{\frac{5}{6}} \sin\big( \frac{5\pi}{4}\big) + \cdots \Big] \,. \nonumber
\]

On the other hand, at $\xi = \infty$ or $\eta = 0$ the theory also becomes integrable, which is known as $E_8$ field theory due to its connection to $E_8$ algebra\footnote{When $m=0$ with $h$ is real and nonvanishing, this IFT is massive integrable with 8 stable particles. Their masses $M_p$ with $p=1,2,\dots, 8$ are proportional to the components of Frobenius vector of $E_8$ Cartan matrix, and the scatterings of stable particles also reflect the properties of $E_8$ root system in various ways \cite{Zamolodchikov:1989fp}.}. We call the point $\xi = \infty$ or $\eta = 0$ the $E_8$ point. Near the $E_8$ point at small $\eta$, IFT can be regarded as the perturbation of $E_8$ field theory with the relevant operator $\varepsilon (x)$, and the scaling functions are expandable as power series in $\eta = {m}/{h^{\frac{8}{15}}}$. For example, at $\eta \to 0^-$ or $u \to +\infty$:
\[
\mathcal M_1(\eta) = \sum_{n=0}^{\infty}M^{(n)}_1 \eta^n\,, \quad \text{and}\quad
\hat M_1(u) = u^{\frac{4}{15}} \sum_{n=0}^{\infty}M^{(n)}_1 u^{-\frac{4n}{15}} \,, \label{M1regularExpansion1}
\]
where the coefficients can be represented using the form factors of $\varepsilon(x)$ in $E_8$ field theory. The exact values of $M^{(0)}_1$ and $M^{(1)}_1$ were computed in the literature \cite{Fateev:1993av}\cite{Delfino:1996xp}\cite{Fateev:1997yg}\cite{Alekseev:2011my}\cite{Xu:2022mmw}. As a result, the behavior of discontinuity $\Im m \, \hat M_1(u+i0)$ at $u \to -\infty$ is given by the continuation of \eqref{M1regularExpansion1}, as the expansion:
\[
\Im m\, \hat M_1(u + i0) = u^{\frac{4}{15}} \sum_{n=0}^{\infty}M^{(n)}_1 u^{-\frac{4n}{15}} \sin \big(\frac{4(1-n)}{15}\big) \,.
\]

\subsection*{Some properties of the low-T regime}
In this subsection, we will briefly introduce the physical interpretations of IFT in the low-T phase. When $T<T_c$ with $h = 0$, the $\mathbb{Z}_2$ symmetry is broken by the nonvanishing spontaneous magnetization. Because of that, the low-T phase of $T < T_c$ is also known as the ordered phase. Ising model become the playground of more abundant interesting phenomena in the low-T phase, like confinement, nucleation and false vacuum decay. Some of them are reflected by the $M_1$ analyticity in low-T, and will be discussed in the following sections.

When $h = 0$, the vacua of low-T IFT are double degenerate. Denote the degenerate vacua as $| \pm \rangle$, they differ by the sign of spontaneous magnetizations, as:
\[
\langle \pm | \sigma (x) | \pm \rangle = \pm \bar \sigma = \pm \bar s |m|^{\frac{1}{8}} \,, \quad \text{with} \quad
\bar s = 2^{\frac{1}{12}} e^{-\frac{1}{8}} A_G^{3/2} = 1.35783834 \dots \,, \label{sigmaVEV}
\]
where $A_G$ is the Glaisher's constant \cite{mccoy2013two}. The degeneracy of $|\pm \rangle$ vacua is lifted once a small nonvanishing $h$ is presented. With small positive $h$, the interaction raises the energy density of $| + \rangle$, and lowers the one of $| - \rangle$. Thus $| - \rangle$ becomes the true stable vacuum, while $| + \rangle$ loses its stability and becomes the metastable vacuum. Similar to many other systems with slightly broken double degenerated vacua, fluctuations lead to the instability of metastable phase, and the decay/tunneling from the metastable vacuum to the true vacuum is described by the theory of nucleation. See \cite{fisher1967theory}\cite{langer2000theory} for systematic discussions.

\begin{figure}[!h]
\centering
\includegraphics[width=0.6\textwidth]{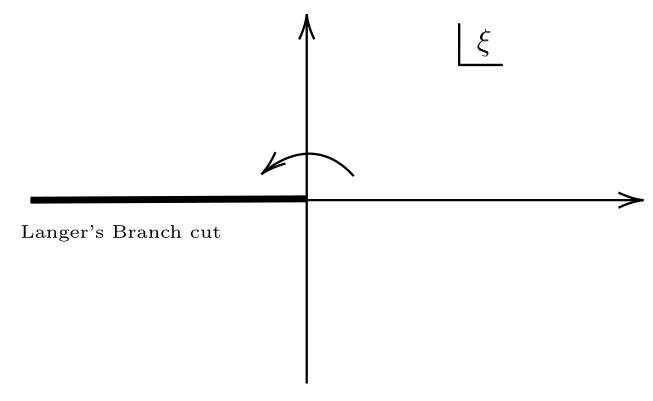}
\caption{Phase diagram of $\mathcal{F}(\xi)$ on the complex $\xi$ plane when $T<T_c$. Point $\xi =0$ is the Fisher-Langer point, as a branching essential singularity. The branch cut extends along the negative real axis of $\xi$, which define the principle branch of $\mathcal F(\xi)$.}
\label{LangerBC}
\end{figure}

In the low-T phase, analyticity of scaling functions are very different comparing to the ones of the high-T phase. Consider the scaling function of free energy density in low-T, which is defined as:
\[
\mathcal F(\xi) = \frac{1}{|m|^2}\big( F(m,h) - \frac{m^2}{8\pi} \log m^2 \big) \,, \label{FScalingFunctionDef}
\]
where the second term is of subtracting Onsager's singularity \cite{fonseca2003ising}. $\mathcal F(\xi)$ is a single-valued continuous function on positive real axis of $\xi$, and is expandable as an asymptotic series in powers of $\xi$ at small positive $\xi$. On the complex $\xi$-plane, the low-T analyticity of $\mathcal F(\xi)$ follows from the Yang-Lee theory of the low-T, see e.g. \cite{fonseca2003ising}\cite{langer2000theory}\cite{andreev1964singularity}\cite{gunther1980goldstone}\cite{lowe1980instantons}\cite{harris1984ising}.

The low-T analyticity of function $\mathcal F(\xi)$ states that $\mathcal F(\xi)$ can be continued to a function of generic complex $\xi$. The function is analytic on the whole $\xi$-plane, except for $\xi = 0$ where a branching essential singularity is located. We call $\xi = 0$ the Fisher-Langer point. In \cite{fonseca2003ising}, the associated Fisher-Langer's branch cut is chosen to be along the negative real $\xi$-axis, extending from $\xi = -\infty$ to $\xi \to 0^-$. This defines the principle branch of $\mathcal F(\xi)$ in the low-T regime. Along the Fisher-Langer's branch cut, $\mathcal F(\xi)$ has a nonvanishing discontinuity. The analytic structure of $\mathcal F(\xi)$ is as shown in Fig.\ref{LangerBC}. No other singularity exist either at other complex $\xi$ or along the negative real $\xi$-axis.

The low-T analyticity is suggesting the associated dispersion relation, as representing the scaling function by an integration along the Fisher-Langer's branch cut. The discontinuity of $\mathcal F(\xi)$ on the branch cut is pure imaginary, as $\text{Disc}\, \mathcal F(\xi) = 2 i \Im m \, \mathcal F(\xi+i0)$ \footnote{At low-T, the relation $\mathcal F(\xi^*) = \mathcal F( \xi)^*$ still holds, and at real negative $\xi$ the discontinuity is $\text{Disc}\, \mathcal F(\xi) = \mathcal F(\xi + i0) - \mathcal F(\xi - i0)=  2i \Im m \, \mathcal F(\xi + i0)$.} with $\xi < 0$. The discontinuity $\text{Disc}\, \mathcal F(\xi)$ is controlled by its behaviours at both $\xi \to -\infty$ and $\xi \to 0^-$. Near the $E_8$ point $\xi \to +\infty$, similar to \eqref{M1regularExpansion1} but with positive $\eta$, $\mathcal F$ can be expanded as integer powers of $\eta$, or equivalently of $\xi^{-\frac{8}{15}}$. The expansion can be safely continued as $\xi \to  e^{\pm \pi i}\xi$ with sufficient large $\xi$, and would result in the expansions of $\Im m \, \mathcal F(\xi + i0)$ at $\xi \to -\infty$.

However, the behaviour of $\mathcal F(\xi)$ become less regular near the essential singularity $\xi = 0$. For small positive $\xi$, $\mathcal F(\xi)$ is expandable in integer power series of $\xi$, as\cite{fonseca2003ising}:
\[
\mathcal F(\xi) = -\bar s \xi + \tilde G_2 \xi +\cdots\,, \label{FLowExpansion}
\]
in the low-T regime. The continuation of \eqref{FLowExpansion} would find vanishing $\Im m \, \mathcal F(\xi+i0)$ on the negative real axis, by using $\xi \to  e^{\pm \pi i}\xi$. Instead, at $\xi \to 0^-$, $\Im m \, \mathcal F(\xi + i0)$ is given by the theory of nucleation \cite{fisher1967theory}\cite{langer2000theory}\cite{Voloshin:1985id}, as:
\[
\Im m\, \mathcal F(\xi + i0) \sim \frac{\lambda}{4\pi} e^{ - \frac{\pi}{\lambda} }\,, \quad \text{where} \quad \lambda = -2\bar s \xi > 0 \,.  \label{FmetaCondensation}
\]
The expression \eqref{FmetaCondensation}, as the imaginary part of metastable free energy density, is related to the decay rate of metastable vacuum. See \cite{langer2000theory} for more details.

In \cite{fonseca2003ising}, numerical approximation of $\Im m \, \mathcal F(\xi + i0)$ was provided, based on the above analysis of $\xi \to -\infty$ and $\xi \to 0^-$ asymptotic expressions. The low-T dispersion relation of $\mathcal F(\xi)$ was verified numerically, supporting the low-T analyticity as in Fig.\ref{LangerBC}. It's believed that other scaling functions exhibit similar analyticity, but the form of essential singularities (as \eqref{FmetaCondensation}) would be different. The low-T analyticity of $\hat M_1(\xi)$ will be analyzed and numerical verified in Sec.\ref{Section2}.

\subsection*{Spectrum in low-T domain}
While any QFT in its Euclidean version can be interpreted as statistical mechanics of fields, the $1+1$d Minkowskian version of IFT \eqref{IFTaction} gives rise to interesting relativistic particle theory. In the low-T domain with small $\xi$, the massive excitations of 1d Ising spin chain are interpreted as Ising mesons with tower-like meson spectrum. This serves as a result of the confining interaction from external magnetic field, and is known as the McCoy-Wu scenario \cite{Wu:1975mw}\cite{mccoy1978two}\cite{mccoy2013two}.

At zero magnetic field $h$, the Ising action \eqref{IFTaction} can be reduced to the free fermion action:
\[
\mathcal A_{\text{FF}} = \frac{1}{2\pi} \int d^2 x \big( \psi \bar\partial \psi + \bar\psi \partial \bar\psi + i m \bar\psi \psi   \big) \,, \label{FFaction}
\]
where $\psi$ and $\bar\psi$ denote the Majorana fermions. When $T< T_c$ and $h=0$, the fermions can be understood as the domain walls separating regions filled with different degenerate vacua. When a small external magnetic field is presented, the vacua degeneracy is lifted, giving rise to a confining interaction between the fermions (which will be referred to as "quarks"). The interaction energy is proportional to the separation between two quarks, and the bound state can be interpreted as one string attaching with two quarks. The string tension is $f_0 = 2 \bar\sigma h$ when $h$ is small, and can be interpreted as difference of energy densities between stable and metastable vacuum. The bounded pairs of quarks would behave as mesons, just like other 2d confining models \cite{tHooft:1974pnl} but without color symmetry structure. As the result, the massive excitations of 2d low-T IFTs are understood roughly as the confined pairs of quarks. Under the 2-quark approximation, various methods can be applied to compute the approximated Ising meson spectrum.

In the non-relativistic limit, the hamiltonian of a single Ising meson involves of two quarks with a linear potential:
\[
H = 2 m + \frac{p_1^2 + p_2^2}{2m} + 2\bar\sigma h |x_1 - x_2| \,. \label{nonSRhamiltonian}
\]
The non-relativistic mass spectrum is given by eigenvalues of \eqref{nonSRhamiltonian}, reads:
\[
M_n \to 2m + m \, \big(\frac{2\bar\sigma h}{m^2}\big)^{\frac{2}{3}} z_n \quad \text{with} \quad \text{Ai}(-z_n) = 0 \,, \label{nonSRspectrum}
\]
where $z_n$ label consecutive zeros of the Airy function $\text{Ai}(-z)$ \cite{Wu:1975mw,mccoy1978two}. This approximation works well for $n \ll m^2/\bar\sigma h$. For higher-levels with $n \sim m^2/\bar\sigma h$, relativistic effects must be included, and what is called the semi-classical quantization is more suitable.

Consider a relativistic Ising meson consisting of two quarks. In the center of mass frame, the classical trajectories of both quarks moves back and forth periodically. The trajectories look like lentils, with each quark draw part of a hyperbola in each period. Parameterizing the rapidities of quarks as $\pm \beta$, the classical trajectories are given by:
\[
p = \frac{p_1 - p_2}{2} = - m \sinh\beta \,, \quad \quad
x = x_1 - x_2 = \frac{m}{\bar\sigma h} (\cosh \vartheta - \cosh \beta)\,,
\]
where $\vartheta$ is the maximum rapidity which occurs at quarks crossing. The action per period of the classical trajectory reads:
\[
\oint p dx = \frac{m^2}{\bar\sigma h} \int_{-\vartheta}^{+\vartheta} \sinh^2\beta d\beta  = \frac{\sinh 2\vartheta - 2\vartheta}{\lambda} \,,
\quad \text{with} \quad \lambda = \frac{2 \bar\sigma h}{m^2}\,,
\]
which should be quantized. Quantization condition is given by the Bohr-Sommerfeld rule, as:
\[
2 \oint p d x = 2\pi (N + \frac{1}{2}) \,,
\]
where $N$ is a positive odd integer due to fermionic nature of quarks. By denoting $N = 2n-1$, the quantization gives discrete $\vartheta_n$ and meson spectrum, as:
\[
\sinh 2\vartheta_n - 2\vartheta_n = 2\pi \lambda (n - \frac{1}{4}) \,,
\quad \quad M_n = 2 m \cosh \vartheta_n\,, \label{WKBspectrum}
\]
with $M_n$ is also called the WKB mass of meson. Higher order correction of \eqref{WKBspectrum} is available, as of the form:
\[
\sinh 2\vartheta_n - 2\vartheta_n = 2\pi \lambda (n - \frac{1}{4}) - \lambda^2 \bar S_1(\theta) + O(\lambda^3) \,,
\quad \quad M_n = 2 m \cosh \vartheta_n\,, \label{WKBspectrum2}
\]
with $\bar S_1(\theta)$ is a meromorphic function of $\theta$, with explicit form given in \cite{Fonseca:2006au}\cite{Rutkevich:2009zz}.

A more systematic analysis of Ising meson spectrum using 2-quark approximation requires solving the corresponding Bethe-Salpeter equation \cite{Fonseca:2006au}. The Bethe-Salpeter equation is derived using the known matrix elements $\langle q(\theta) q(-\theta) | \sigma(0) | q(\theta') q(-\theta') \rangle$. The details of \cite{Fonseca:2006au} is too lengthy to be introduced here, but we would collect the useful conclusions. At small positive $\lambda$, the Ising meson masses admit the expansion (see also \cite{Rutkevich:2009zz}):
\[
M_n(m,h) = 2m + m \sum_{k=1}^{\infty} c_n^{(k)} \lambda^{\frac{2k}{3}} + m \sum_{l=4}^{\infty} {d}_n^{(l)} \lambda^{\frac{2l+1}{3}} \,. \label{MesonSeriesExpansion}
\]
Roughly speaking, $M_n$ can be expanded as an asymptotic series of $\lambda^{\frac{2}{3}}$, plus corrections of power series of $\lambda^{\frac{1}{3}}$ starting from relatively higher orders. Some coefficients of \eqref{MesonSeriesExpansion} were computed in \cite{Fonseca:2006au}, while the non-relativistic spectrum \eqref{nonSRspectrum} and WKB spectrum \eqref{WKBspectrum} approximate well the leading coefficients $c_n^{(k)}$ in \eqref{MesonSeriesExpansion}.

Although all the above methods give good approximations of the Ising meson spectrum, they cannot explain the decay of heavier mesons and the inelastic scattering processes, because these phenomena include multi-quark effects. For any meson with mass $M_n >  2 M_1$, the decay channel into two lightest meson opens, and the $n$-th meson is no longer stable. The 2-quark approximations become less accurate when $M_n$ approaches $2M_1$. We would instead denote the stable mesons as $M_p$ with $p = 1,2,3,\cdots, P$. The number of stable mesons $P$ decreases with growing $\xi$. In the strong coupling limit $\xi \to +\infty$ or $\eta \to 0^+$, the Ising field theory can be regarded as perturbed $E_8$ field theory. At sufficiently large $\xi$, only $M_{1,2,3}$ are below the stability threshold $2 M_1$, and other higher mesons cross their stability thresholds at finite values of $\xi$'s. The analyticity of higher meson masses is an interesting questions, and we plan to address it elsewhere.

\subsection*{Remarks}
\begin{figure}[!h]
\centering
\includegraphics[width=0.7\textwidth]{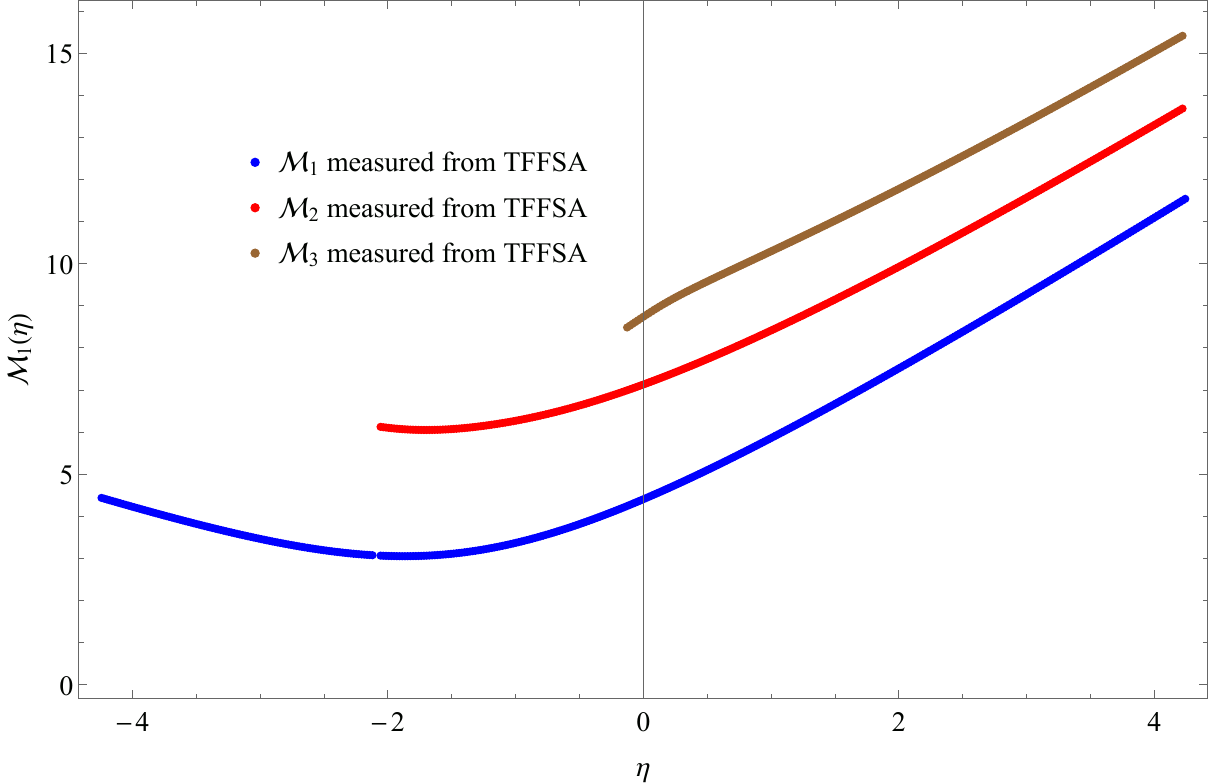}
\caption{The spectrum of scaling function $\mathcal M_p(\eta)$ with $p=1,2,3$ near $\eta = 0$, which represent the masses of three lightest particle. The numerical data was obtained using the truncated free fermion space approach (TFFSA). We are using the same set of data as in \cite{zamolodchikov2013ising} and \cite{Xu:2022mmw}.}
\label{M123_eta_TFFSA}
\end{figure}

Similar to the previous works \cite{fonseca2003ising,zamolodchikov2013ising,Xu:2022mmw,Xu:2023nke}, this paper is a combination of both analytic computations and numerical analysis. With the same manner, the numerical data of Ising spectrum which are used in this work was obtained via Truncated Free Fermion Space Approach (TFFSA), and the details of TFFSA were well discussed in \cite{fonseca2003ising} and \cite{Xu:2022mmw}. For a given $\xi$ (or $\eta$), TFFSA is able to numerically compute the energy levels $E_n(R)$ of IFT on a cylinder with good accuracy\footnote{Usually 5 to 6 digits when not close to any critical point.} (here we denote circumference of the cylinder as $R$).
At large $R$, the energy gaps $\Delta E_n(R) = E_n(R)- E_0(R)$ approach some constants $M_n$ exponentially, which are identified as the energies of particle states. In this work, we use the same set of $\mathcal M_p(\eta)$ or $\hat{M}_p $ data as in \cite{zamolodchikov2013ising} and \cite{Xu:2022mmw}. The behaviour of $\mathcal M_{1,2,3}$ near $\eta = 0$ are as plotted in Fig.\ref{M123_eta_TFFSA}, and the details of data measurements are omitted here.

In Sec.\ref{Section2}, we propose the low-T analyticity conjecture of the stable particle masses $M_{1,2,3}$, with the associated low-T dispersion relations are given in \eqref{M123_LowT_disperion_eq}. The low-T analyticities are checked in this section numerically, as is shown in Fig.\ref{M123_LowT_disperion1} and \eqref{M123_LowT_disp_integrals_checks}. In Sec.\ref{Section3}, we turn to the complex $\eta$-plane and propose the extended analyticity conjecture of the lowest mass $M_1$. The extended dispersion relation is given as \eqref{M1EDispequation}, and the extended analyticity of $M_1$ is numerically verified in various ways. Finally, in Appendix \ref{Appendix_A}, we give a detailed analysis on the "instanton-like" terms, which come from the tunneling between the degeneracy-lifted vacua. These "instanton-like" terms are small enough to be negligible in the discussion of Sec.\ref{Section2} and Sec.\ref{Section3}. In Appendix \ref{Appendix_B}, we cover some technical details of constructing the discontinuities for the extended dispersion relation \eqref{M1EDispequation}.

\section{Analyticity of ${\hat M}_p(\xi),\,p=1,2,3$ and the Fisher-Langer's branch cut}\label{Section2}

In the low-T phase with $T<T_c$ (or $m>0$), IFT has at least three stable particles, with their masses denoted as $M_p,\,p=1,2,3$. Here we focus on the analytic properties of the associated scaling functions $\hat{M}_p(\xi) = M_p(m,h) / |m|$ on the complex $\xi$-plane. For any $p \ge 4$, the corresponding Ising meson becomes unstable at sufficiently large $\xi$ (or small positive $\eta$), making its analyticity more complicated. For the stable mesons $p = {1,2,3}$ one expects that the functions $\hat M_p(\xi)$, being analytically continued from the positive part of the real $\xi$-axis, are analytic on the whole complex $\xi$-plane, except at the Fisher-Langer essential singularity at $\xi = 0$, and the associated Fisher-Langer branch cut; the latter extends along the negative part of the real axis, from $-\infty$ to $0$. Under this assumption the $M_p(\xi)$ can be expressed through the discontinuity across this branch cut via the corresponding dispersion integral.

\subsection*{Analytic continuation of $M_p$}

At small $\xi$ the functions ${\hat M}_p (\xi)$ admit the asymptotic expansion \eqref{MesonSeriesExpansion}, with $\lambda = 2{\bar s}\,\xi$. On the other hand, when $\xi \gg 1$ the IFT can be regarded as $E_8$ field theory perturbed by the energy density operator $\varepsilon(x)$, and the scaling functions ${\cal M}_p(\eta)$ enjoy regular expansions
\[
\mathcal M_p(\eta) = \frac{M_p(m,h)}{|h|^{{8}/{15}}} = M^{(0)}_p +  M^{(1)}_p \eta +  M^{(2)}_p \eta^2 +   M^{(3)}_p \eta^3 + \cdots\,, \label{Mp_Eta_expansion1}
\]
which converge at sufficiently small $\eta$. Equivalently, on the complex $\xi$-plane the expansion \eqref{Mp_Eta_expansion1} reads
\[
\hat M_p(\xi) = \frac{M_p(m,h)}{|m|} = \xi^{\frac{8}{15}}\Big(M^{(0)}_p +  M^{(1)}_p \xi^{-\frac{8}{15}} +  M^{(2)}_p \xi^{- \frac{16}{15}} +   M^{(3)}_p \xi^{-\frac{8}{5}} + \cdots\Big)\,. \label{Mp_expansion_mn}
\]
Some of the leading coefficients of \eqref{Mp_Eta_expansion1} can be computed via the form-factor perturbation theory \cite{Fateev:1993av}\cite{Delfino:1996xp}\cite{Fateev:1997yg}\cite{Alekseev:2011my}\cite{Xu:2022mmw}. For example, $M_1^{(0)}$ is available from the mass-coupling relation of $E_8$ field theory, and $M_1^{(1)}$ was computed using the 2-point form-factor of $\varepsilon(x)$ operator \cite{Delfino:1996xp}. The coefficients $M_2^{(0)}$ and $M_3^{(0)}$  follow straightforwardly from $E_8$ mass ratios \cite{Zamolodchikov:1989fp},
\[
{M_2^{(0)}}/{M_1^{(0)}} = 2 \cos \frac{\pi}{5} \,, \quad \text{and} \quad
{M_3^{(0)}}/{M_1^{(0)}} = 2 \cos \frac{\pi}{30} \,.
\]
The coefficients of $M_2^{(1)}$ and $M_3^{(1)}$ can also be computed by using the form factors of the $E_8$ field theory \cite{Delfino:1996xp}. For higher coefficients, only numerical results are available. We list some of them in Tab.\ref{Tab:Mp_coefficients}, as well as numerical values of exact results from \cite{Delfino:1996xp}.

As is shown in Fig.\ref{LangerBC}, the Fisher-Langer branch cuts of $\hat M_p(\xi)$ extends along the real negative $\xi$-axis. The discontinuities of ${\hat M}_p(\xi)$ across the branch cut are pure imaginary,
\begin{equation}
\text{Disc}\,  \hat M_p(x): = \hat M_p(x + i0) - \hat M_p(x - i0) = 2 i \Im m \, \hat M_p(x + i0) \,, \label{FLdisc}
\end{equation}
where $x = -\xi > 0$. At sufficiently large positive $x$ they enjoy the convergent expansions
\[
\Im m \, \hat M_p(x + i0) = x^{\frac{8}{15}} \sum_{n = 0}^{\infty} M_p^{(n)} \sin \big[ \frac{8\pi(1 - n)}{15} \big] x^{-\frac{8 n }{15}} \,, \label{Im_Mp_expansion_mn}
\]
which follows from the term-by-term continuation of \eqref{Mp_expansion_mn}.

On the other hand, at sufficiently small $\xi$ the same discontinuity \eqref{FLdisc} admits asymptotic expansion
\begin{gather}
\text{Disc}\,{\hat M}_p(x) =  2i\,\Theta(x)\, \Big (\sum_{k=1}^{\infty} {\hat c}_p^{(k)} x^{\frac{2k}{3}}  + \sum_{l=4}^{\infty} {\hat d}_p^{(l)} x^{\frac{2l+1}{3}} \Big) \,, \label{MesonSeriesExpansion2}
\end{gather}
which follows from the continuation of \eqref{MesonSeriesExpansion}, with $\Theta(x)$ is the step function. Here the coefficients are:
$${\hat c}_p^{(k)} = (2{\bar s})^\frac{2k}{3} c_p^{(k)}\sin(\frac{2\pi k}{3})\,, \quad \text{and} \quad \hat{d}_p^{(l)} = (2\bar s)^\frac{2l+1}{3} d_p^{(l)} \sin \big( \frac{\pi (2l+1)}{3} \big) \,.$$
While few terms of this expansions approximate well the desired discontinuities only at very small $x$, at larger $x$ the approximative power of these expansions is expected to deteriorate rapidly, not to mention that only limited number of the terms are available explicitly \cite{Fonseca:2006au}\cite{Rutkevich:2009zz}. Instead, the leading semiclassical approximation \eqref{WKBspectrum} gives reasonably accurate results for ${\hat M}_p(\xi)$ at positive real $\xi$, for all
but very large $\xi \lessapprox 2.0 $ \cite{Fonseca:2006au}. Thus, we will make an assumption (confirmed by the result of this Section) that analytic continuation of \eqref{WKBspectrum} to complex values of $\lambda$, including the negative part of the real axis, gives comparably accurate
approximation of $\hat M_p(\xi)$ at all complex $\xi$ with $|\text{arg}(\xi)| \leq \pi$, as long as $|\xi| \lesssim 0.5$.


Below I will build an approximation for $\Im m\,{\hat M}_p(x)$, by combining the analytic continuation of the leading semiclassical spectrum \eqref{WKBspectrum} with the partial sums of the series \eqref{Im_Mp_expansion_mn}. For every $p=1,2,3$, the four leading terms of \eqref{Im_Mp_expansion_mn} match well with the analytic continuation of \eqref{WKBspectrum} at $|\xi| \simeq x_p^\text{Int}$,
\[
x^{\text{Int.}}_1 \approx 0.369927 \,,  \quad
x^{\text{Int.}}_2 \approx 0.469173 \,,  \quad
x^{\text{Int.}}_3 \approx 0.247037 \,.  \quad\label{xpIntValues}
\]
Then, we simply approximate $\text{Disc}\,{\hat M}_p(-\xi)$ with the four terms truncation of \eqref{Im_Mp_expansion_mn} when $\xi \le -x_p^\text{Int}$, while using the analytic continuation of \eqref{WKBspectrum} for $ 0> \xi > -x_p^\text{Int}$. The approximated discontinuities for $p = 1,2,3$ are as shown in Fig.\ref{M123_LowT_discontinuities_piecewise1}.

\begin{figure}[!bh]
\centering
\includegraphics[width=0.7\textwidth]{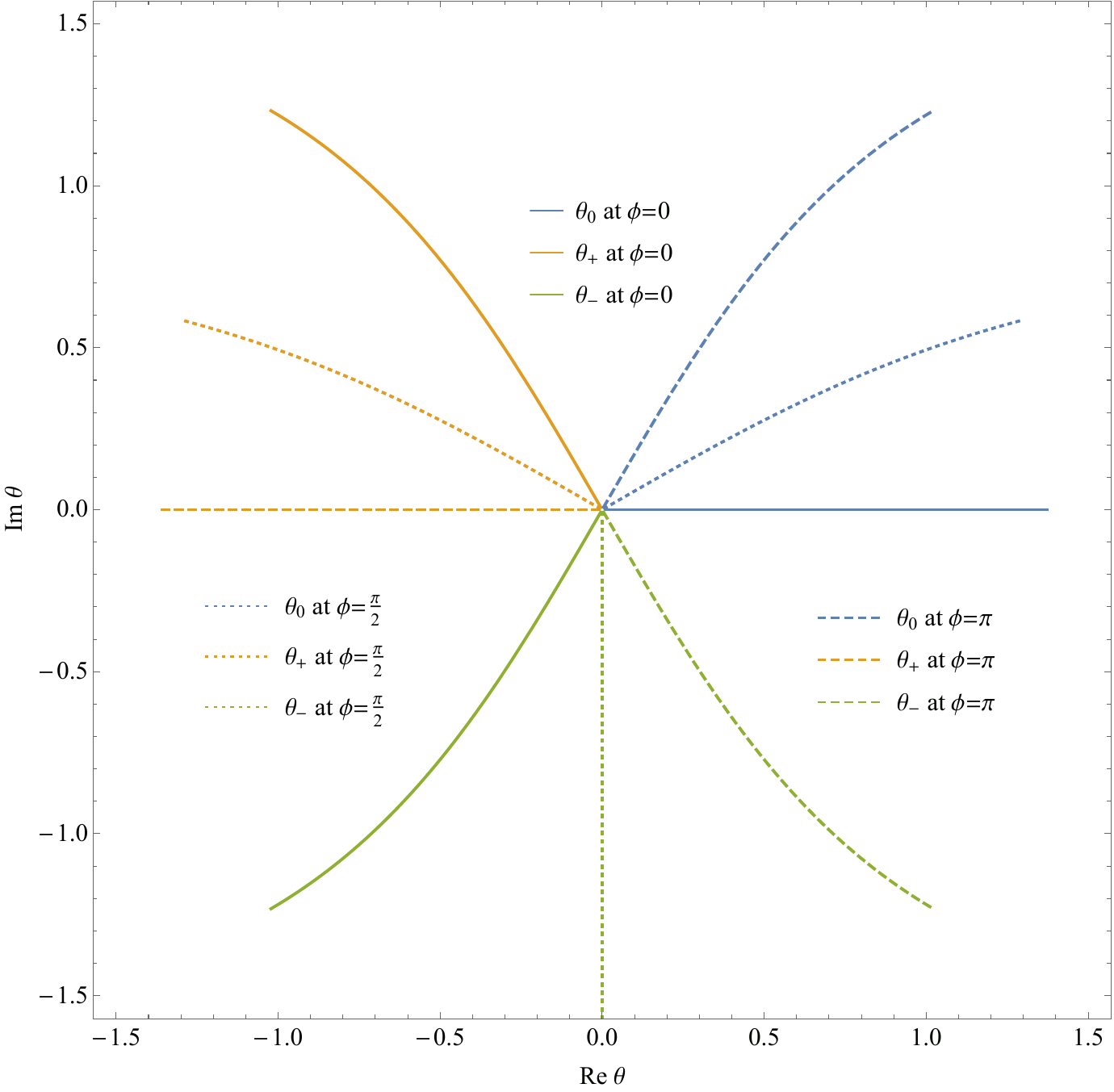}
\caption{The evolution of solutions to the transcendental equation \eqref{thetatdefinitioneq}, along with the phase angle $\phi = \text{Arg}\, t$ varies from $0$ to $\pi$. Three solutions of \eqref{thetatdefinitioneq}, namely $\theta_0$, $\theta_+$ and $\theta_-$, are given by curves in different colors. Each of them rotates counterclockwisely near the origin.}
\label{RotatingSCSolutions1}
\end{figure}

Here we shall explain how the analytic continuation of \eqref{WKBspectrum} is obtained. The quantization condition \eqref{WKBspectrum} has the form
\[
\sinh 2\theta(t) - 2\theta(t) = t \label{thetatdefinitioneq}\,,
\]
where we have consider $t$, once as discrete values in \eqref{WKBspectrum}, now as a continuous parameter, and study the behavior of the solution $\theta(t)$. At given real and positive $t$, there exist three solutions, one is real (denoted as $\theta_0(t)$) and two are complex (denoted as $\theta_+(t)$ and $\theta_-(t)$).  At real $t$ the two complex solutions $\theta_+(t)$ and $\theta_-(t)$ are complex conjugate to each other. It is the real solution $\theta_0(t)$ which is taken in the quantization of the meson mass spectrum, but it is instructive to follow all the three families. At small $t$ all three solutions behave as the three cubic root of $t$, while at large $t$ the pair of solutions $\theta_+(t)$ and $\theta_-(t)$ monotonically approach $-\infty \pm \frac{\pi}{2} i$, as shown in Fig.\ref{RotatingSCSolutions1}.

Now, let's make the parameter $t$ complex, as writing $t \to e^{i \phi} |t|$ with some phase $\phi$. When the phase $\phi$ increases the pattern of solutions changes as follows. At small $t$ it just rotates by the angle $\phi/3$, while at large $t$ the asymptotics of solutions $\theta_0(t)$ and $\theta_\pm$ generally changes. More precisely, when $\phi$ increases from zero to $\pi/2$, the solution $\Re e\,\theta_0(t) \to +\infty$ and $\Re e \, \theta_\pm(t) \to -\infty$ as $|t| \to \infty$, while their imaginary parts monotonically increase/decrease. At $\phi=\pi/2$ the solution $\theta_-(t)$ becomes pure imaginary at all $|t|$, extending along the negative part of imaginary axis. The two other solutions behave as $\theta_+ = -\Re e \theta_0 + i \Im m \, \theta_0$, and are as shown in Fig.\ref{RotatingSCSolutions1}. When $\phi$ further increases from $\pi/2$ to $\pi$ the solutions keep "rotating" counterclockwise, while preserving the signs of their real parts. Until at $\phi=+\pi$ the solution $\theta_+(t)$ becomes real negative, and $\theta_-(t)$ becomes complex conjugate of $\theta_0(t)$. Therefore, at $\xi < -x^{\text{Int}}_p$ we take
\begin{gather}
\text{Disc} \, {\hat M}_p (\xi) := {\hat M}_p(\xi+i0) - {\hat M}_p(\xi-i0) = 2i\,\Im m\,{\hat M}_p(\xi+i0) \nonumber\\
\approx \quad \Theta(-\xi)\,\left[2\cosh\theta_+(- t)-2\cosh\theta_-(- t)\right] = 4 i \Theta(-\xi)\, \Im m \, \big( \cosh\theta_+(- t) \big)  \,, \label{semixdisc}
\end{gather}
where $\Theta(x)$ is the step function, and $t$ in the last expression is to be set to $4\pi \bar s \,\xi\,(p-1/4)$. This provides an
leading semiclassical approximation for the discontinuity, which is expected to work well at sufficiently small $|\xi|$.

It is fair to note that the above approximation ignores possible "instanton-like" contributions to \eqref{FLdisc}, which are
exponentially small at $x << 1$ but may become significant at larger $x$. However, as will be argued in Appendix \ref{Appendix_A}, these
contributions are numerically very small as compared to the above approximation, and disregarding them is legitimate.

\subsection*{The low-T dispersion relations for $M_1$, $M_2$ and $M_3$.}

The low-T analyticity conjecture for $M_1$, $M_2$ and $M_3$ states that: the associated scaling functions ${\hat M}_1(\xi)$,
${\hat M}_2(\xi)$, ${\hat M}_3(\xi)$ are analytic in the whole complex $\xi$-pane, with the Fisher-Langer branch cut extending along the real axis from $-\infty$ to $0$, as is depicted in Fig.\ref{LangerBC}. The discontinuities $\text{Disc}\,\hat M_p(-x) = 2i \Im m \,\hat M_p(-x + i0)$ are pure imaginary continuous functions at $x>0$. Under this conjecture the following dispersion relations hold,
\[
\hat M_p(\xi) = 2 + \xi \int_{0}^{+ \infty} \frac{dx}{\pi} \frac{\Im m \,\hat M_p(-x + i0)}{x(x+\xi)} \,, \label{M123_LowT_disperion_eq}
\]
for all complex $\xi$ in Fig.\ref{LangerBC}. For the function $\mathcal M_p(\eta) = \eta \, \hat M_p(\xi  = \eta^{{15}/{8}})$ these relations take the form
\[
\mathcal{M}_p  (\eta) = 2\eta + \frac{15 \eta}{8\pi} \int_0^{\infty} dz \, \frac{z^{{7}/{8}} \Im m \,\hat M_p(- z^{-15/8} + i0)}{z^{{15}/{8}} + \eta^{{15}/{8}}} \,. \label{M123_LowT_disperion_eq_eta}
\]
Under the above low-T analyticity conjecture, this dispersion relation holds as long as $|\text{arg} \eta| \leq 8\pi/15$.
Below the certain approximations for $\Im m \,\hat M_p(-x + i0)$ are proposed, and the dispersion relation
\eqref{M123_LowT_disperion_eq_eta} are verified numerically.

\begin{table}[h]
\begin{center}
 \begin{tabular}{|c|c|c|c|}
 \hline
 $M^{(0)}_1$ & $M^{(1)}_1$ & $M^{(2)}_1$ & $M^{(3)}_1$ \\
 \hline
 $4.405$ & $1.295$ & $0.2002$ & $-0.051$ \\
  \hline
 $M^{(0)}_2$ & $M^{(1)}_2$ & $M^{(2)}_2$ & $M^{(3)}_2$ \\
 \hline
 $7.130$ & $1.113$ & $0.2116$ & $-0.040$ \\
 \hline
  $M^{(0)}_3$ & $M^{(1)}_3$ & $M^{(2)}_3$ & $M^{(3)}_3$ \\
 \hline
 $8.740$ & $1.978$ & $-0.3804$ & $-1.0$ \\
 \hline
 \end{tabular}
\caption{Coefficients of $\mathcal M_p(\eta)$ expansions \eqref{Mp_Eta_expansion1} near $\eta = 0$. Among the coefficients $M_p^{(k)}$, $M^{(0)}_p$ \& $M^{(1)}_p$ were computed using form factor perturbation theory \cite{Fateev:1993av}\cite{Delfino:1996xp}\cite{Fateev:1997yg}\cite{Alekseev:2011my}\cite{Xu:2022mmw}, while $M^{(2)}_p$ and $M^{(3)}_p$ were measured from the spectrum data \cite{zamolodchikov2013ising}\cite{Xu:2022mmw}\cite{Xu:2023nke}.}
\label{Tab:Mp_coefficients}
\end{center}
\end{table}

\begin{figure}[!th]
\centering
\includegraphics[width=0.9\textwidth]{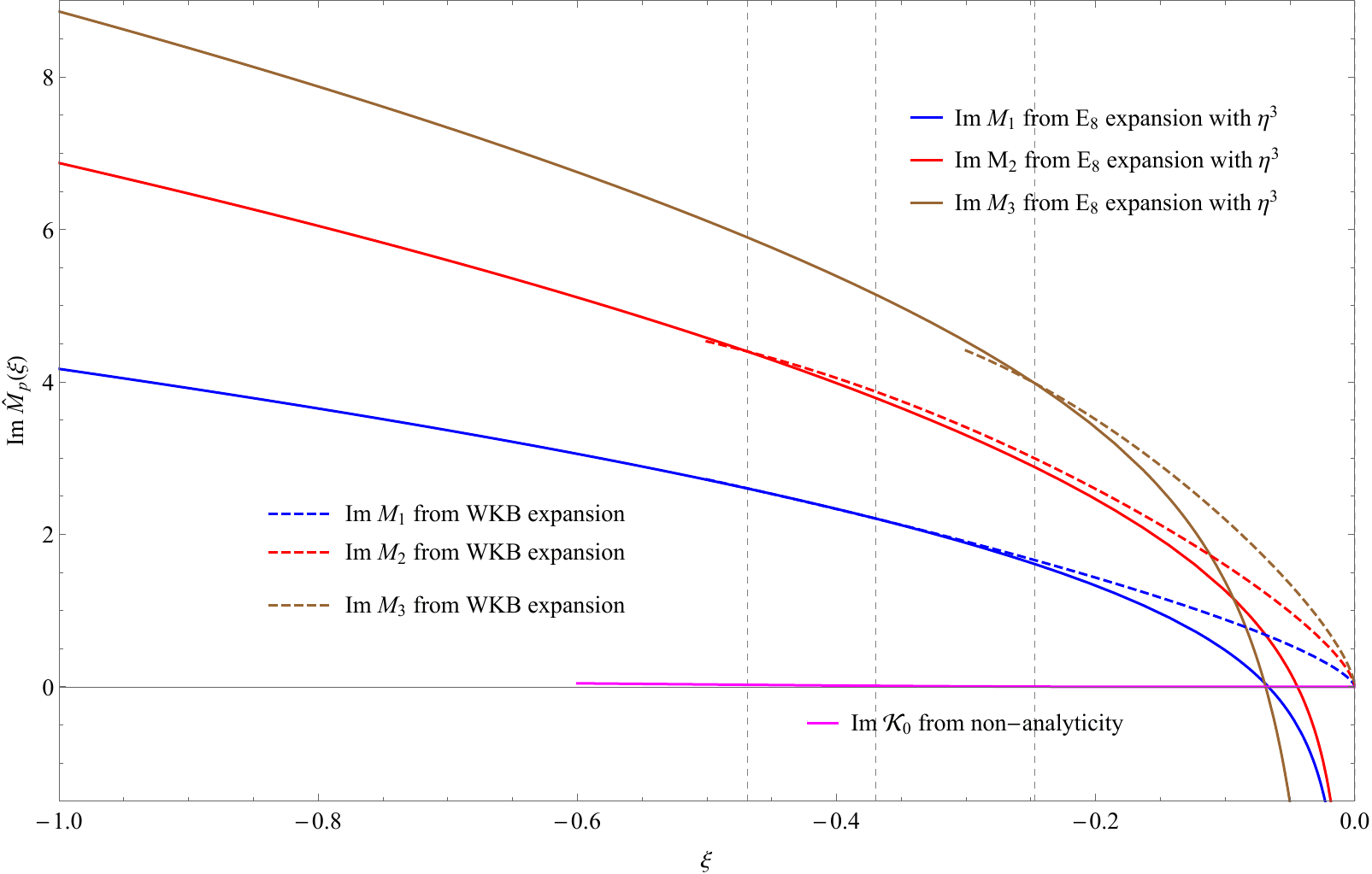}
\caption{Matching the functions $\Im m \, \hat M_p(\xi +i0)$ for $\xi < 0$, as illustrate the behaviours of the discontinuities \eqref{FLdisc} along the Fisher-Langer's branch cut. In this plot, blue, red and brown solid curves are showing $\Im m \, \hat M_p(\xi +i0)$ with $ p =1,2,3$ following the expansions \eqref{Im_Mp_expansion_mn}, which works for $\xi \to -\infty$. The coefficients $M_p^{(k)}$ of \eqref{Im_Mp_expansion_mn} are those of Tab.\ref{Tab:Mp_coefficients}, and the series are truncated till the terms with $M_p^{(3)}$. On the other hand, blue, red and brown dashed curves are illustrating the continuation of semiclassical quantization spectrum, see \eqref{semixdisc} and the related discussion. The curves of the same color match at different $\xi = - x_p^{\text{Int}}$, with the values given in \eqref{xpIntValues}. Lastly, the magenta curve is the instanton-like term $\Im m \, \mathcal K_0$ \eqref{ImKappa0Expression}, which is almost negligible comparing to other contributions, thus omitted in the dispersion integral \eqref{M123_LowT_disperion_eq}.
}
\label{M123_LowT_discontinuities_piecewise1}
\end{figure}

At large $\xi$, $\Im m \,\hat M_p(-x + i0)$ can be approximated by few leading terms of the expansion
\eqref{Im_Mp_expansion_mn}. Two leading coefficients of this expansion are known exactly from form factor perturbation theory
\cite{Delfino:1996xp}, and two more can be extracted from the TFFSA for $\mathcal M_p(\eta)$. These coefficients are
presented in Table \ref{Tab:Mp_coefficients}. On the other hand, at small $|\xi|$ the leading semiclassical formula \eqref{semixdisc} provides good approximation. Simultaneous plots of four leading terms of the expansion \eqref{Im_Mp_expansion_mn} and the semiclassical approximation \eqref{semixdisc} are presented in Fig.\ref{M123_LowT_discontinuities_piecewise1} for $p=1,2,3$. The plot of the estimate \eqref{Kappa0Expression} of the "instanton-like" contribution is also shown in this Figure, where one can see that the latter contribution is negligible. One can also observe that the above large $\xi$ and small $\xi$ approximations match at some intermediate values of $\xi = - x_p^{\text{Int}}$, shown above in \eqref{xpIntValues}. Therefore, for each $p=1,2,3$, we propose the approximation

\begin{figure}[!thb]
\centering
\includegraphics[width=0.9\textwidth]{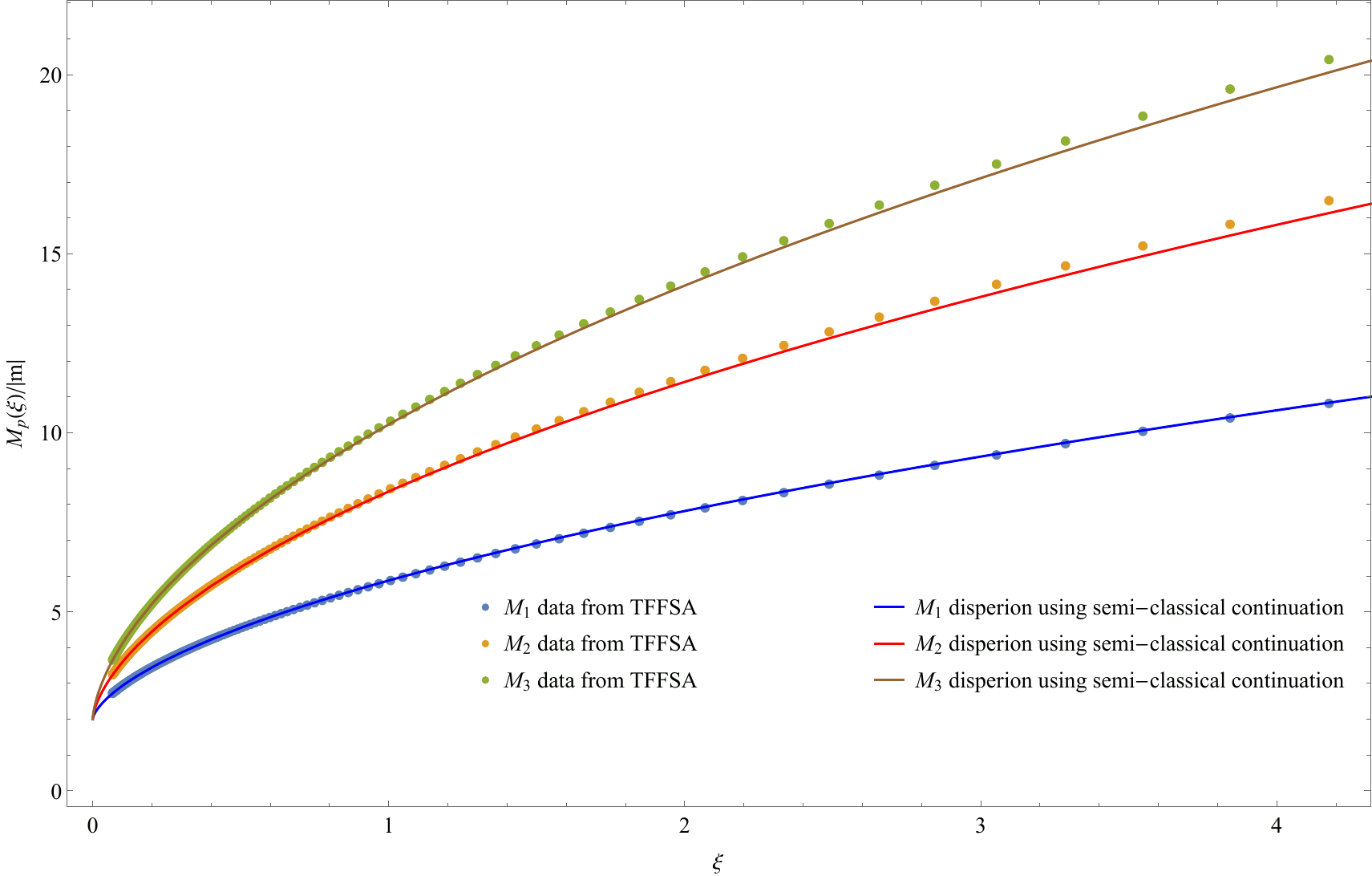}
\caption{The verifications low-T dispersion relations \eqref{M123_LowT_disperion_eq} with $ p =1,2,3$ by numerical integrations. The discontinuities \eqref{FLdisc} of the integration are approximated using \eqref{DiscApprox}, namely using \eqref{Im_Mp_expansion_mn} for $\xi < - x_p^{\text{Int}}$ and \eqref{semixdisc} for $\xi > - x_p^{\text{Int}}$. The results of numerical integrations are as solid curves, which are to be compared to the mass spectrum from TFFSA, as in Fig.\ref{M123_eta_TFFSA}, with transformed from $\mathcal M_p(\eta)$ to $\hat M_p(\xi)$ \eqref{M1etaRealDef}. All three levels are showing good match between the dispersion integrals and measured data, supporting the low-T analyticity conjecture.}
\label{M123_LowT_disperion1}
\end{figure}

\[
\text{Disc} \, {\hat M}_p^\text{approx}(-x) = \bigg\{\nfrac{\text{Eq}.\eqref{MesonSeriesExpansion2}\ \text{truncated to}\ n=3\ \text{for} \ x>x^{\text{Int.}}_p}
{\text{Eq}.\eqref{semixdisc}\ \qquad \text{for}\ x<x^{\text{Int.}}_p}\,.\label{DiscApprox}
\]
With this approximation the dispersion integral in \eqref{M123_LowT_disperion_eq_eta} can be evaluated numerically, and compared to the ${\hat M}_p(\xi)$ obtained from the TFFSA data. This comparison is shown in Fig.\ref{M123_LowT_disperion1}. The close match supports our low-T analyticity conjecture.

Another independent verification can be made as follows. One notes that the approximation \eqref{DiscApprox} does not involve any information of the coefficients $M_p^{(1)}$, since in \eqref{Im_Mp_expansion_mn} the term $\sim x^{-8/15}$ enters with vanishing coefficient. The dispersion relation allows one to restore this coefficient. Expanding \eqref{M123_LowT_disperion_eq_eta} at large $\eta$ one obtains
\[
M^{(1)}_p = 2 + \frac{15 }{8\pi} \int_0^{\infty} dz \, \Big[ \frac{\Im m \,\hat M_p(- z^{-15/8} + i0)}{z} -\frac{ M^{(0)}_p \sin(\frac{8\pi}{15}) }{z^2}\Big] \,, \label{Mp1DiscCheck}
\]
Numerical evaluation of this integral with the approximation \eqref{DiscApprox}, we have
\[
M^{(1)}_1 \Big|_{\text{disp}} = 1.262 \,, \quad
M^{(1)}_2 \Big|_{\text{disp}} = 1.084 \,, \quad
M^{(1)}_3 \Big|_{\text{disp}} = 1.541 \,, \label{M123_LowT_disp_integrals_checks}
\]
which are to be compared with the exact values in Tab.\ref{Tab:Mp_coefficients}. One can see that these values deviate from the exact coefficients by $3\%$ for $M_1^{(1)}$ and $M_2^{(1)}$ to $30\%$ for $M_3^{(1)}$. This results are at least compatible with the low-T analyticity conjecture, but also show that our approximation for the discontinuity \eqref{DiscApprox} while qualitatively correct but needs further refinement. We leave this question for further analysis.

\section{Extended dispersion relation of $M_1$}\label{Section3}
The analyticity properties of scaling function $\hat M_1(\xi)$ in high-T regime and low-T regime have been discussed separately in \cite{Xu:2022mmw} and Sec.\ref{Section2}, and in this section, we will sew both together by considering the analyticity in terms of the scaling parameter $\eta$. We will turn to the analyticity of the function $\mathcal M_1(\eta)$, which is defined as \eqref{M1etaRealDef} below. Specifically, we propose "$M_1$ extended analyticity" in this section, and support it by verifying numerically the associated dispersion relation.

\subsection*{$\mathcal M_1(\eta)$ on the complex plane of $\eta$}

At real $m$ and positive $h$, $\mathcal M_1(\eta)$ is defined as follows:
\[
\mathcal M_1(\eta) = \frac{M_1(m,h)}{h^{\frac{8}{15}}}  =
\begin{cases}
- \eta \, \hat M_1 \big(\xi^2 = (-\eta)^{-\frac{15}{4}} \big) \quad & \text{High-T:} \quad \eta < 0  \\
+ \eta \, \hat M_1 \big(\xi = \eta^{-\frac{15}{8}} \big) \quad & \text{Low-T:} \quad \eta > 0
\end{cases} \,,\label{M1etaRealDef}
\]
which is the lightest mass $M_1$ measured in the unit of $h^{\frac{8}{15}}$. \eqref{M1etaRealDef} also illustrated how $\mathcal M_1(\eta)$ is related to the functions $\hat M_1$ of high-T and low-T regimes, differs by the sign of $\eta$. Furthermore, when $ m < 0 $ ($T > T_c$) and $h$ is a pure imaginary, $\eta$ becomes complex with $\text{Arg}\, \eta = \pm \frac{11 \pi}{15}$, and $\mathcal M_1(\eta)$ becomes a complex function. By denoting $\eta = -y e^{\mp \frac{4\pi i}{15}}$, the scaling function $\hat{\mathcal M_1 }(y) = |\mathcal M_1(\eta)|$ with $y>0$ is related to $\hat M_1(\xi^2)$ of high-T by:
\[
\hat{\mathcal M_1 }(y) =
 \big|\mathcal M_1 \big( \eta = -y e^{\mp \frac{4\pi i}{15}} \big)\big| = e^{\pm \frac{4\pi i}{15}} \mathcal M_1 \big( \eta = -y e^{\mp \frac{4\pi i}{15}} \big) = y \, \hat M_1 \big( \xi^2 = - y^{-\frac{15}{4}}  \big)\,. \label{M1yImaginaryDef}
\]
In Fig.\ref{PhaseDiagramXi}, $\hat M_1(\xi^2)$ is a single-valued function for $ -\xi^2_0 \le \xi^2 < 0 $, thus \eqref{M1yImaginaryDef} works for $y \ge Y_0$. As was discussed in \cite{Xu:2022mmw} and Sec.\ref{Section2}, $\mathcal M_1(\eta)$ was measured at real $\eta$ and real $y$ with good accuracy\footnote{For any real $\eta$ or generic $y > Y_0$, the fitting procedure in \cite{Xu:2022mmw} yields accuracy of 3 to 5 significant digits of $\hat M_1$. When $y$ becomes very close to $Y_0$ the Yang-Lee point, due to the large correlation length, the accuracy of $\hat M_1$ falls into 2 digits.}, by using the finite size spectrum from TFFSA (see e.g. \cite{fonseca2006ising} and \cite{Xu:2022mmw}). These numerical data will be used in the following analysis, see e.g. Fig.\ref{M1EdispPlotCorrect1} and Fig.\ref{M1EdispMunitPlotCorrect1}.

\begin{figure}[!th]
\centering
\subfigure[]{\includegraphics[width=7.0cm]{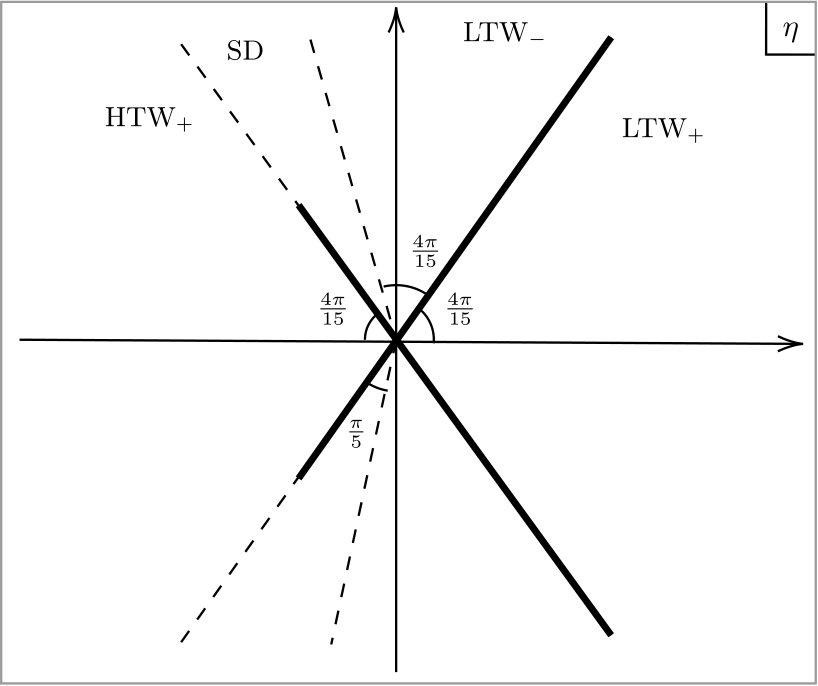}}
\hspace{1em}
\subfigure[]{\includegraphics[width=7.0cm]{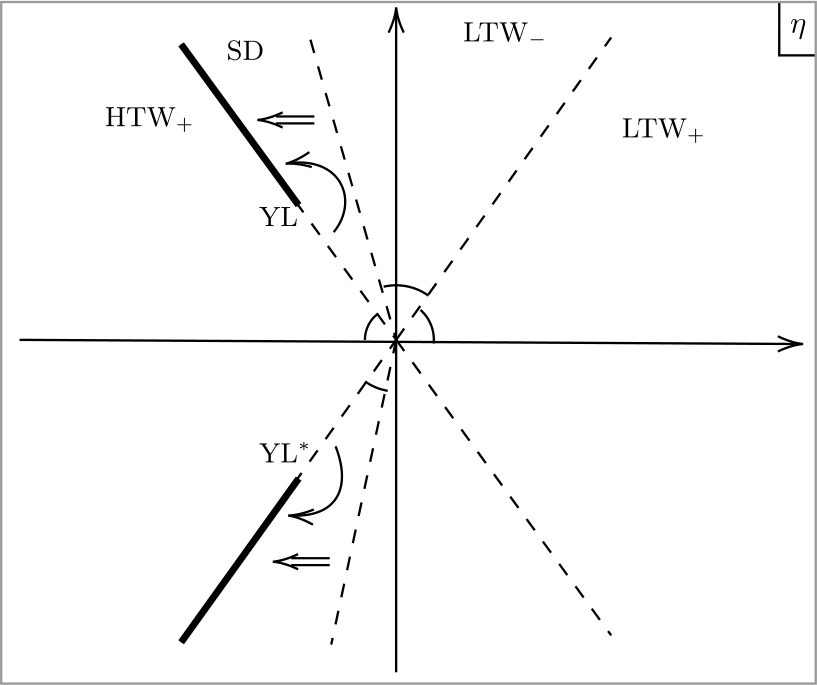}}
\caption{The analyticity structure of $\mathcal M_1(\eta)$  on the complex $\eta$-plane, with different choices of Yang-Lee branch cut and primary branches of $\mathcal M_1(\eta)$. $|\text{Arg}\,\eta| \le \frac{8\pi}{15}$ is the low-T wedge (LTW), $|\text{Arg}\, (- \eta)| \le \frac{4\pi}{15}$ is the high-T wedge (HTW) and $\frac{8\pi}{15}\le |\text{Arg}\,\eta| \le \frac{11\pi}{15}$ is the shadow domain (SD). The edges of Fisher-Langer's branch cut become the rays $\text{Arg}\,\eta = \pm \frac{8\pi}{15}$. (a): Mapping the high-T complex $u = \xi^2$ plane Fig.\ref{PhaseDiagramXi} to the complex $\eta$-plane with $\eta = -u^{-4/18}$. Along the ray $\eta = -y e^{\mp \frac{4\pi i}{15}}$, the Yang-Lee branch cut extends from $y = Y_0$ to $y = -\infty$. (b): After the rotation of the Yang-Lee branch cut, it becomes extending from $y = Y_0$ to $y = +\infty$. This defines the principle branch of $\mathcal M_1(\eta)$. Now LTW and SD are compatible with HTW on the same complex $\eta$ plane. The statement of extended analyticity is regarding analytic structure of SD, as there exist no singularities within the shadow domain or along the rotated Yang-Lee branch cut.}
\label{eta_planes}
\end{figure}

At generic complex $\eta$, $\mathcal M_1(\eta)$ can be reached by continuing \eqref{M1etaRealDef} from the real $\eta$-axis. On the complex $\eta$-plane, the first sheets of function $\hat M_1(\xi)$ for low-T and $\hat M_1(\xi^2)$ of high-T are mapped to different wedges \cite{fonseca2003ising}. The $\xi$-plane of $\hat M_1(\xi)$ for low-T, as was illustrated in Fig.\ref{LangerBC}, is mapped to the wedge $ |\text{Arg}\,\eta | < \frac{8}{15}\pi $. We call this wedge the low-T wedge (LTW). The boundaries of the low-T wedge are the rays $\text{Arg}\,\eta  = \pm \frac{8}{15}\pi$, which are the images of upper/lower edges of the Fisher-Langer's branch cut (see Fig.\ref{LangerBC}). LTW is further decomposed into the two domains, as $\text{LTW}_+$ for $|\text{Arg}\,\eta | < \frac{4}{15}\pi $, which was mapped from $\Re e \, \xi > 0$, and $\text{LTW}_-$ for $\frac{4}{15}\pi  < |\text{Arg}\,\eta | < \frac{8}{15}\pi $, which was mapped from $\Re e \, \xi < 0$. The LTW is as shown on the right half $\eta$-plane of Fig.\ref{eta_planes}(a) and Fig.\ref{eta_planes}(b).

When $T >T_c$, the image of complex $\xi^2$-plane (see Fig.\ref{PhaseDiagramXi}) is the wedge $|\text{Arg}\,(-\eta) | < \frac{4}{15}\pi $, which is called the high-T wedge (HTW). The function $\mathcal M_1(\eta)$ in the HTW was mapped from $\hat M_1(\xi^2)$ on the first sheet of complex $\xi^2$-plane, as is shown in Fig.\ref{eta_planes}(a). The negative real axis of $\xi^2$-plane represents IFT with pure imaginary magnetic field, and its upper and lower edges mapped to the rays $\text{Arg}\,\eta = \pm \frac{11}{15} \pi$. On the complex $\eta$-plane, the Yang-Lee singularities are located at $ \eta = - Y_0 e^{\mp \frac{4\pi i}{15}}$, which serve as branching points of scaling functions. As is shown in Fig.\ref{eta_planes}(a), the choice of Yang-Lee branch cut in Fig.\ref{PhaseDiagramXi} is mapped to $ y \le Y_0$ on the complex $\eta$-plane, along the rays $\eta = -y e^{\mp \frac{4\pi i}{15}} $.

However, the mapping of Fig.\ref{eta_planes}(a) separates high-T wedge and low-T wedge in different isolated domains. From low-T point of view, the high-T $\xi^2$-plane of Fig.\ref{PhaseDiagramXi} is mapped to certain domain under the branch cuts in Fig.\ref{eta_planes}(a). It can be exposed by rotating the branch cuts as shown in Fig.\ref{eta_planes}(b). From high-T wedge point of view, the rotation of branch cuts is equivalent to redefine $\mathcal M_1(\eta)$ for $|\text{Arg}\,\eta | < \frac{11}{15}\pi $, in order to match the function $\mathcal M_1(\eta)$ of the low-T wedge. The function $\mathcal M_1(\eta)$ in the high-T wedge remains intact. Both LTW (including $\text{LTW}_+$ \& $\text{LTW}_-$) and HTW can now be embedded on the same $\eta$-plane, as shown in Fig.\ref{eta_planes}(b). However, the complex $\eta$-plane is not fully covered by these wedges. The uncovered domain, namely $\frac{8}{15}\pi  < |\text{Arg}\,\eta | < \frac{11}{15}\pi$, is called shadow domain (SD). From the low-T point of view, the shadow domain is on the second sheet of Fig.\ref{LangerBC}, as behind the Fisher-Langer's branch cut. The shadow domain does not correspond to any physically well defined system.

After the rotation of branch cuts, the function $\mathcal M_1(\eta)$ becomes single-valued function on the rays $\text{Arg}\, \eta = \frac{11}{15}\pi$, for $ 0 < y \le Y_0$. \eqref{M1yImaginaryDef} now becomes:
\begin{equation}
\mathcal M_1 \big( \eta = -y e^{\mp \frac{4\pi i}{15}} \big) = y e^{\mp \frac{4\pi i}{15}}  \, \hat M_1 \big( \xi^2 = - y^{-\frac{15}{4}} \pm i0  \big)\,. \label{Mcal1WrongSideDef}
\end{equation}
For $y \ge Y_0$, the discontinuities of scaling functions across the Yang-Lee branch cut of Fig.\ref{eta_planes}(b) are different. On the lower edge $\eta = -y e^{\mp (\frac{4\pi i}{15}+i0)}$, the function $\mathcal M_1(\eta)$ follows its behaviours in the HTW. While differently on the upper edge $\eta = -y e^{\mp (\frac{4\pi i}{15}-i0)}$, the function $\mathcal M_1(\eta)$ sits on the edges of the shadow domain, and can be reached by the continuation from the low-T wedge. The continuation corresponds to entering the second sheet of Fig.\ref{LangerBC}, which is behind the Fisher-Langer's branch cut. Certain analytical structure in the shadow domain would affect the continuation. As a result, the analyticity of function $\mathcal M_1(\eta)$ on the complex $\eta$-plane is completely determined by its discontinuities along the rotated Yang-Lee branch cut, and the analytic structures in the shadow domain, as illustrated in Fig.\ref{eta_planes}(b).

\subsection*{The $M_1$ extended analyticity conjecture, and its dispersion relation}
The minimal assumption of $\mathcal M_1(\eta)$ analyticity is similar to the extended analyticity conjecture in \cite{fonseca2003ising}, which states that on the first sheet of $\eta$-plane (see Fig.\ref{eta_planes}(b)), the only singularities of the scaling function are the Yang-Lee branching points. In other words, no other singularities exist either within the shadow domain or on the rotated Yang-Lee branch cut. The Yang-Lee branching points are the closest non-trivial singularities behind the Fisher-Langer's branch cut. We call this analyticity statement of $\mathcal M_1(\eta)$ as $M_1$ extended analyticity conjecture. On the rotated Yang-Lee branch cut, the discontinuity of $\mathcal M_1(\eta)$ is a smooth function on $ Y_0 < y < +\infty$, and determines the behaviour of $\mathcal M_1(\eta)$ on the $\eta$-plane via $M_1$ extended dispersion integral, which will be formulated as follows.

At generic complex $\eta$ on Fig.\ref{eta_planes}(b), the contour integration reproducing $\mathcal M_1 (\eta)$ (with two terms subtracted) reads:
\begin{gather}
\mathcal M_1(\eta) - \Big( M_1^{(0)} + M_1^{(1)}\eta \Big) = \eta^2 \oint_{\mathcal C_\eta} \frac{d\eta'}{2\pi i} \frac{\mathcal M_1(\eta') -  M_1^{(0)} - M_1^{(1)}\eta'}{\eta'^2(\eta'-\eta)} \label{M1EDispequation00} \\
= 
\eta^2 \int_{Y_0}^{+\infty} \frac{dy}{2\pi i} \Big(\frac{\mathcal M_1 (y e^{+ \frac{11 \pi i}{15} + i0}) - \mathcal M_1 (y e^{+ \frac{11 \pi i}{15} - i0})}{y^2 e^{+ \frac{11}{15}\pi i} (y e^{+ \frac{11}{15}\pi i} - \eta)} 
+\frac{\mathcal M_1 (y e^{- \frac{11 \pi i}{15} + i0}) - \mathcal M_1 (y e^{- \frac{11 \pi i}{15} - i0})}{y^2 e^{- \frac{11}{15}\pi i} (y e^{- \frac{11}{15}\pi i} - \eta)}
\Big)
\,, \label{M1EDispequation0}
\end{gather}
where $\mathcal C_\eta$ denotes a small contour circling counterclockwise around $\eta$. In \eqref{M1EDispequation0}, $\mathcal C_\eta$ is deformed into the integration from $\eta' = e^{\frac{11 \pi i}{15}} Y_0$ to $ \eta' = e^{\frac{11 \pi i}{15}} (+\infty)$ with its complex conjugation\footnote{By $\mathcal M_1(\eta)$ is a real function when $\eta$ is real, at complex $\eta$ $\mathcal M_1(\eta^*) = \big(\mathcal M_1(\eta)\big)^*$.}. The coefficients $M_1^{(0)}$ and $M_1^{(1)}$ in \eqref{M1EDispequation00} are known exactly \cite{Delfino:1996xp}\cite{Xu:2022mmw}, and the numerical values were given in Tab.\ref{Tab:Mp_coefficients}.

Now we discuss the discontinuity on the numerator of \eqref{M1EDispequation0}. After rotating the branch cuts, \eqref{M1yImaginaryDef} gives $\mathcal M_1(\eta)$ on the lower edge of Yang-Lee branch cut at $\eta = y e^{\pm (\frac{11}{15}\pi i + i0) }$, which sits on the high-T wedge side. On the rays $\text{Arg} \, \eta = \pm \frac{11}{15}\pi$, $\mathcal M_1\big(\eta = -y e^{\pm \frac{4 \pi i}{15}} \big)$ is a single-valued function for $y \le Y_0$, and for $y > Y_0$ we denote:
\begin{gather}
\mathcal M_1 (y e^{+ \frac{11 \pi i}{15} \pm i0}) = e^{+ \frac{11 \pi i}{15}} \bar{\mathcal M}_1(y \pm i0) \,, \nonumber\\
\mathcal M_1 (y e^{- \frac{11 \pi i}{15} \mp i0}) = e^{- \frac{11 \pi i}{15}} \bar{\mathcal M}_1(y \mp i0) \,, \label{M1barDef}
\end{gather}
as the definition for $\bar{\mathcal M}_1(y \pm i0)$. Unlike in Fig.\ref{eta_planes}(a), $\bar{\mathcal M}_1(y \pm i0)$ are not complex conjugation of each other, and the discontinuity of the extended dispersion relation is defined as:
\[
{\Delta}_1(y) = \frac{1}{2 i }\Big[ \bar{\mathcal M}_1(y+ i0) - \bar{\mathcal M}_1(y- i0) \Big] \,, \label{Delta1Definition}
\]
which is a complex function for $y \ge Y_0$. With the help of \eqref{Delta1Definition}, \eqref{M1EDispequation0} is reduced as:
\begin{gather}
\mathcal M_1 (\eta) = M_1^{(0)} + M_1^{(1)} \eta + \frac{2 \eta^2}{\pi} \int_{Y_0}^\infty \frac{dy}{y^2} \frac{y\, \Re e \, \big( e^{-\frac{11 \pi i}{15}} \Delta_1(y)\big) - \eta\, \Re e \, \Delta_1(y)}{y^2 - 2\cos\big(\frac{11 \pi}{15}\big)\eta y + \eta^2} \,. \label{M1EDispequation}
\end{gather}
We call \eqref{M1EDispequation} the $M_1$ extended dispersion relation. If the $M_1$ extended analyticity conjecture is correct, \eqref{M1EDispequation} should work for any complex $\eta$ on Fig.\ref{eta_planes}(b), including the ones in shadow domain.

Comparing to the extended dispersion relation of free energy density (see eq.(4.8) in \cite{fonseca2003ising}), \eqref{M1EDispequation} is simpler. The Onsager's singularity and other subtractions are absent in the $M_1$ extended dispersion relation \eqref{M1EDispequation}. The integration \eqref{M1EDispequation} converges, because $\lim\limits_{y\to +\infty} \Re e \, \Delta_1(y) = 0$, and no further subtraction is needed.

\subsection*{On the discontinuity $\Delta_1(y)$}
With the conjecture of $M_1$ extended analyticity, there exist no singularity on the rotated Yang-Lee branch cut, and the discontinuity ${\Delta}_1(y)$ is a continuous complex function on the interval $ Y_0 \le y < +\infty$. The approximation of ${\Delta}_1(y)$ can be achieved by using its behaviours near $y = Y_0$ and $y =+\infty$, namely the Yang-Lee singular expansion \eqref{M1SingularExpansion1}, and the expansions at $\eta = \infty$ (equivalently $\xi = 0$, see \eqref{Mp_expansion_mn} and \eqref{MesonSeriesExpansion}). Here we shall establish the approximation of $\Delta_1(y)$ as follows.

In the close vicinity of $y=Y_0$, $\Delta_1(y)$ is controlled by the singular expansions based on Yang-Lee criticality. Before the rotation of branch cuts (see Fig.\ref{eta_planes}(a)), the function $\hat{\mathcal M}_1(y)$, which was defined in \eqref{M1yImaginaryDef}, is single-valued at $y>Y_0$, and admits the singular expansion:
\[
\hat{\mathcal M}_1(y) = (y-Y_0)^{\frac{5}{12}}\big[ \tilde{b_0} + \tilde{c_0} (y-Y_0)^{\frac{5}{6}} + \tilde{b_1}  (y-Y_0) + \cdots \big] \,, \label{M1Singular}
\]
which works in close vicinity of the Yang-Lee point $y=Y_0$. Coefficients in \eqref{M1Singular} read \cite{Xu:2022mmw}\cite{BazhanovYL}:
\[
\tilde b_0 = 3.0754 \,, \quad
\tilde b_1 = 0.8932 \,, \quad
\tilde c_0 = -0.9412 \,. \label{b0b1c0values1}
\]
The singular expansion \eqref{M1Singular} illustrates two branches of $\hat{\mathcal M}_1(y)$ on the both edges of Yang-Lee branch cut in Fig.\ref{eta_planes}(a), which reads (with $y<Y_0$):
\[
\hat{\mathcal M}_1(y \pm i0) = (Y_0 - y)^{\frac{5}{12}}\big[ \tilde{b_0} e^{\pm\frac{5\pi i}{12}} + \tilde{c_0} (Y_0 - y)^{\frac{5}{6}} e^{\pm\frac{5\pi i}{4}} + \tilde{b_1}  (Y_0 - y) e^{\pm\frac{17\pi i}{12}} + \cdots \big] \,, \label{M1SingularBranches}
\]
The corresponding discontinuity in Fig.\ref{eta_planes}(a), before the rotation of the Yang-Lee branch cut, can be expanded as:
\begin{gather}
\hat{\Delta}_1(y) = \frac{1}{2 i }\Big[ \hat{\mathcal M}_1(y+ i0) - \hat{\mathcal M}_1(y- i0) \Big] \label{HatDelta1Definition} \\
= \Theta(Y_0 - y)\, (Y_0 - y)^{\frac{5}{12}}\big[ \tilde{b_0} \sin({\frac{5\pi}{12}}) + \tilde{c_0} (Y_0 - y)^{\frac{5}{6}} \sin({\frac{5\pi }{4}}) + \tilde{b_1}  (Y_0 - y) \sin({\frac{17\pi }{12}}) + \cdots \big] \,, \nonumber
\end{gather}
which is a real function for $y < Y_0$. The rotation of Yang-Lee branch cut, as from Fig.\ref{eta_planes}(a) to Fig.\ref{eta_planes}(b), suggests that the singular expansion representing $\Delta_1(y)$ is a straightforward continuation of \eqref{HatDelta1Definition}, as
\begin{gather}
\Delta_1(y) = \hat{\Delta}_1\big(Y_0 - (y-Y_0)e^{i \pi}\big)  \label{Delta1SingularExpansion1}\\
\simeq \Theta(y - Y_0)\,(y-Y_0)^{\frac{5}{12}} \Big\{  \tilde{b_0} \sin(\frac{5\pi}{12}) e^{\frac{5\pi i}{12}} + \tilde{c_0} \sin(\frac{5\pi}{4}) e^{\frac{5\pi i}{4}}(y-Y_0)^{\frac{5}{6}} + \tilde{b_1} \sin(\frac{17\pi}{12}) e^{\frac{17\pi i}{12}} (y-Y_0) +\cdots \Big\}\,, \nonumber
\end{gather}
which holds for $y>Y_0$ in a close vicinity of the Yang-Lee point. Unlike \eqref{HatDelta1Definition}, $\Delta_1(y)$ is a complex function, with both $\Re e \, \Delta_1(y)$ and $\Im m \, \Delta_1(y)$ contributing to the integral \eqref{M1EDispequation}.

Next we switch to the behaviour of $\Delta_1(y)$ at $y \to +\infty$. At $|\eta| \to +\infty$, from \eqref{M1etaRealDef}:
\[
\mathcal M_1(\eta) \xrightarrow{|\eta| \to \infty}
\begin{cases}
2 \eta & | \text{Arg}\,\eta | < \frac{11}{15}\pi \\
-\eta & | \text{Arg}\,\eta | > \frac{11}{15}\pi
\end{cases} \,,
\]
which suggests $\Delta_1(y) \sim \frac{3}{2}i y$ at $y \to +\infty$. The subleading behaviours at large $y$ follow the different expansions of $\hat M_1$ in high-T and low-T regimes at small $\xi$. When $T < T_c$, $\hat M_1(\xi)$ can be expanded in the series of $\xi^{\frac{2}{3}}$ plus corrections (see \eqref{MesonSeriesExpansion}). Differently when $T > T_c$, $\hat M_1(\xi^2)$ is expanded in integer powers of $\xi^2$ due to the unbroken $\mathbb Z_2$ symmetry. As a result, $\mathcal M_1(\eta)$ behaves as:
\begin{gather}
\mathcal M_1(\eta) = \eta \, \big( 2 + \sum_{k=1}^{\infty} a_1^{(k)} \eta^{-\frac{5}{4}k} + \sum_{l = 4}^{\infty} b^{(l)}_1 \eta^{-\frac{5}{8}(2l+1)} \big) \quad \text{for} \quad | \text{Arg}\,\eta | < \frac{11}{15}\pi  \,, \label{M1etaExpansionLowT1}\\
\mathcal M_1(\eta) = \eta \, \big( -1 + \sum_{k=1}^{\infty} \mu_{2k} \eta^{-\frac{15}{4}k} \big) \quad \text{for} \quad| \text{Arg}\,\eta | > \frac{11}{15}\pi \,, \label{M1etaExpansionHighT1}
\end{gather}
at large $|\eta|$, where the coefficients in \eqref{M1etaExpansionLowT1} are related to the ones in \eqref{MesonSeriesExpansion} via
\[
a_1^{(k)} = (2\bar s)^{\frac{2k}{3}} c_1^{(k)} \,, \quad b_1^{(l)} = (2\bar s)^{\frac{2l+1}{3}} d_1^{(l)} \,.
\]
Since $b_1^{(l)}$ corrections enter $\mathcal M_1(\eta)$ with at least $\eta^{-\frac{37}{8}}$, we will ignore them in the following analysis. Some values of leading $a_k = a_1^{(k)}$ are available in Tab.\ref{Tab:Mp_coefficients}, while various coefficients $\mu_{2k}$ were measured or computed by high-T dispersion relation (see \cite{Xu:2022mmw}), as
\[
\mu_2 \simeq  10.7620 \,, \quad
\mu_4 \simeq -97.22 \,, \quad
\mu_6 \simeq 1396 \,.
\]
Based on the expansion \eqref{M1etaExpansionLowT1} and \eqref{M1etaExpansionHighT1}, the discontinuity $\Delta_1(y)$ at $y \to +\infty$ behaves as:
\begin{equation}
\Delta_1(y) = y \Big( \frac{3}{2}i + \frac{i}{2} \sum_{k=1}^\infty a_k y^{-\frac{5}{4} k } e^{-\frac{11}{12}k\pi i}
- \frac{i}{2} \sum_{k=1}^\infty \mu_{2k} y^{-\frac{15}{4} k } e^{-\frac{11}{4}k\pi i} \Big) \,, \label{Delta1AsymptoticExpansion1}
\end{equation}
However, \eqref{Delta1AsymptoticExpansion1} can be only understood as an asymptotic expansion, namely with zero radius of convergence. Based on the analyticity shown in Fig.\ref{PhaseDiagramXi}, \eqref{M1etaExpansionHighT1} is convergent for $|\eta| > Y_0$ or $|\xi|< \xi_0$. However, the other expansion \eqref{M1etaExpansionLowT1} can only be understood as asymptotic at complex $\eta$, due to the existence of instanton-like corrections (see Appendix \ref{Appendix_A}). Numerically speaking, similar to the case of low-T analysis, the instanton-like terms are negligible, and will be ignored in the following numerical analysis.

\begin{figure}[!th]
\centering
\includegraphics[width=0.8\textwidth]{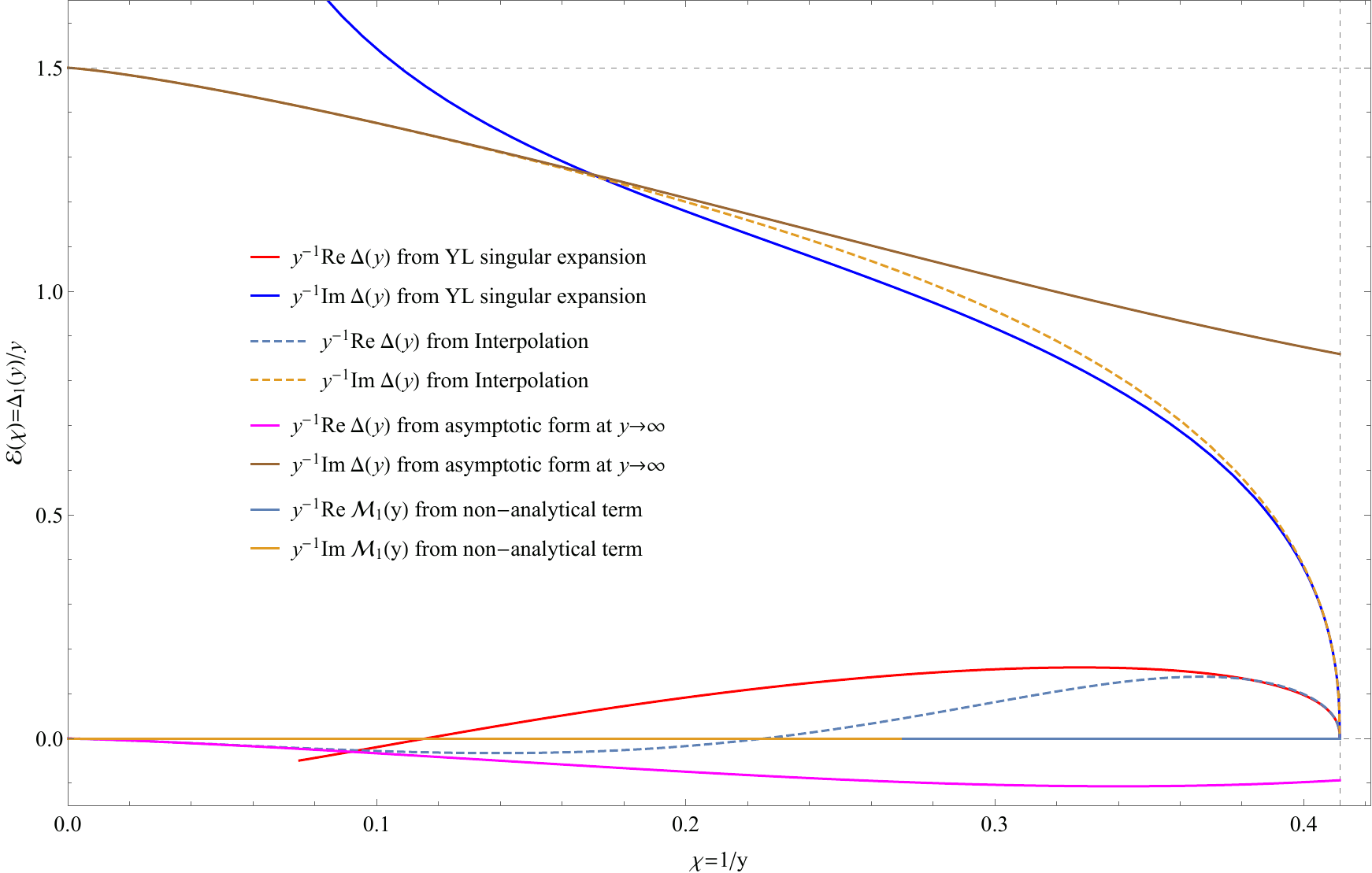}
\caption{The plot illustrating the real and imaginary parts of function $\mathcal E_1(\chi) = \chi \, \Delta_1(y = 1/\chi)$. The horizontal axis is $\chi = 1/y$ on the interval $[0,\mathcal X_0]$, with the Yang-Lee point is located at $\mathcal X_0 = 1/Y_0$. Singular expansion \eqref{Delta1SingularExpansion1} was plotted in red and blue solid lines, as real and imaginary parts. The expansion \eqref{Delta1AsymptoticExpansion1} at $y=\infty$ or $\chi = 0$ was plotted in magenta and brown solid lines. The dashed curves are the interpolations of $\mathcal E_1(\chi)$ between \eqref{Delta1SingularExpansion1} and \eqref{Delta1AsymptoticExpansion1}, see \eqref{E1tSingularExtended} and \eqref{ExtraSingularCoefficientsValues} in Appendix \ref{Appendix_B}. The dashed curves serve as good approximations for the real and imaginary part of $\Delta_1(y)$ along the rotated Yang-Lee branch cut. Finally, the two solid lines which almost overlapping with horizontal axis represents the contribution of instanton-like terms. They are as continuing $z \to y e^{\frac{\pi}{5}}$ in \eqref{HatKappaDef}, see Appendix \ref{Appendix_A} for more details. Their contribution to the dispersion integral \eqref{M1EDispequation} is negligible.}
\label{Continued_Disc_Delta1}
\end{figure}

To approximate $\Delta_1(y)$ on the interval $Y_0\le y < +\infty$, interpolation can be used by matching both the expansions \eqref{Delta1AsymptoticExpansion1} and \eqref{Delta1SingularExpansion1}. The details of interpolation are presented in Appendix \ref{Appendix_B}. As a result of interpolation, the real and imaginary parts of discontinuities are as shown in Fig.\ref{Continued_Disc_Delta1}, where we have defined $\chi = 1/y$ and plot the function $\mathcal E_1(\chi) = \chi \, \Delta_1(y = 1/\chi)$ on the interval $0 \le \chi \le \mathcal X_0 = 1/Y_0$. The imaginary part $\Im m \, \mathcal E_1(\chi)$ stays positive on the interval, with $\Im m \, \mathcal E_1(\chi = 0) = \frac{3}{2}$. The real part $\Re e \, \mathcal E_1(\chi)$ changes from negative to positive when $\chi$ increases from $0$ to $\mathcal X_0$. With the approximation of $\Delta_1(y)$ as in Fig.\ref{Continued_Disc_Delta1}, numerical verifications of $M_1$ extended dispersion relation can be achieved.

\subsection*{Check of $M_1$ extended dispersion relation}

\begin{figure}[!th]
\centering
\includegraphics[width=0.8\textwidth]{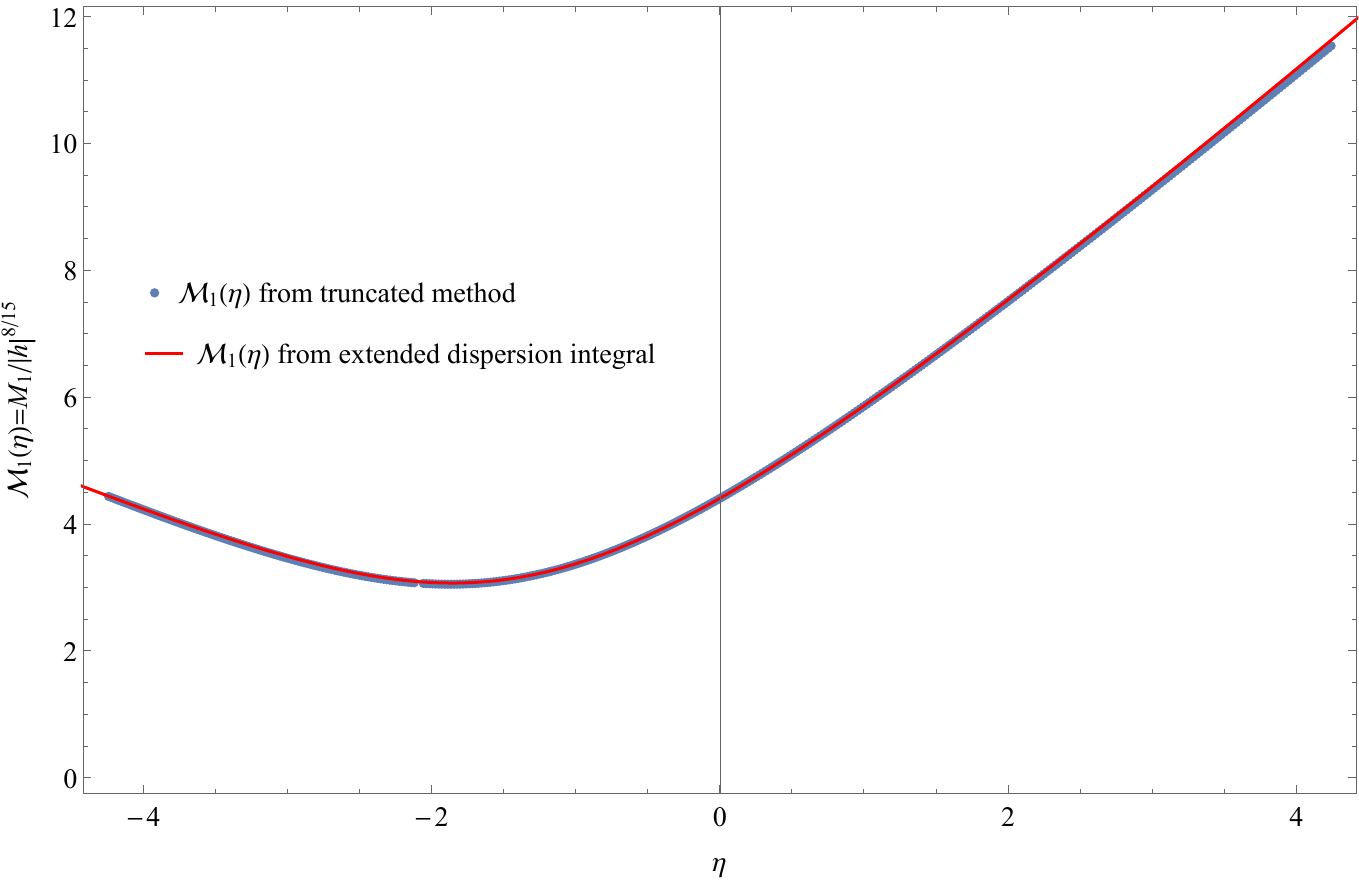}
\caption{A direct verification of $M_1$ extended dispersion relation, by comparing the function $\mathcal M_1(\eta)$ via dispersion relation \eqref{M1EDispequation} and the truncated data (see Fig.\ref{M123_eta_TFFSA}). Numerical data of $\mathcal M_1(\eta)$ are as shown in bullets, which were measured using TFFSA with 5 to 6 digits accuracy. The dispersion integral \eqref{M1EDispequation} is given by red solid curve, with using $\Delta_1(y)$ approximated by the interpolation in Appendix \ref{Appendix_B}, see \eqref{E1tSingularExtended}, \eqref{ExtraSingularCoefficientsValues} and Fig.\ref{Continued_Disc_Delta1}.}
\label{M1EdispPlotCorrect1}
\end{figure}

\begin{figure}[!th]
\centering
\includegraphics[width=0.8\textwidth]{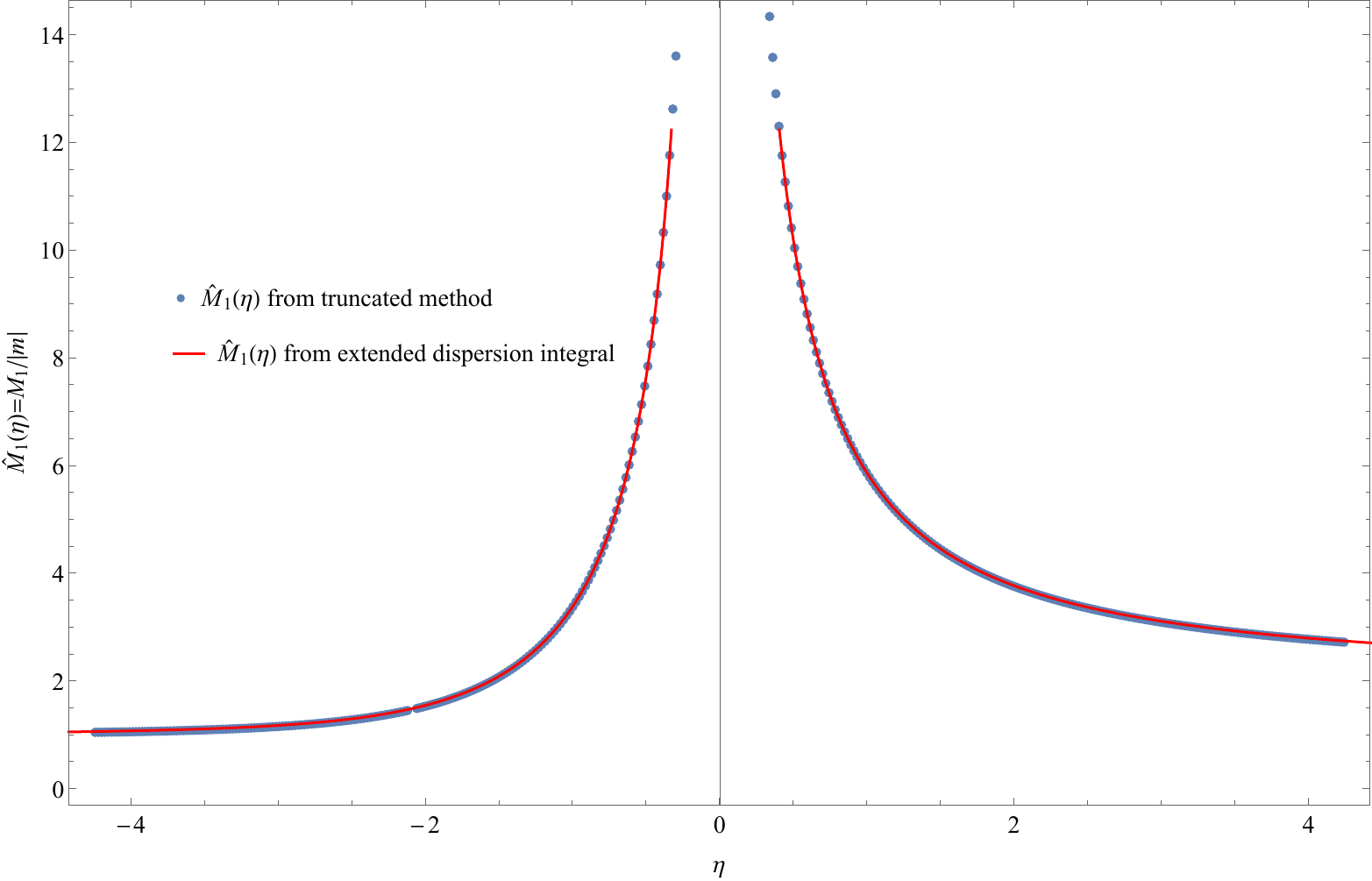}
\caption{Similar to Fig.\ref{M1EdispPlotCorrect1}, a direct comparison of function $\hat{M}_1 = \mathcal M_1 (\eta) /|\eta|$ between the dispersion integral of \eqref{M1EDispequation} and the numerical data. The vertical axis is now $\hat M_1$, as mass of the lightest particle measured in unit of $|m|$. Numerical data from TFFSA are in bullets, and the dispersion integration \eqref{M1EDispequation} is represented in red solid curve, using discontinuities from the interpolation of Appendix \ref{Appendix_B}. The horizontal axis is $\eta$, which is related to $\xi$ via $\eta = - \xi^{-\frac{8}{15}}$ in the high-T regime, and $\eta = \xi^{-\frac{8}{15}}$ in the low-T regime.}
\label{M1EdispMunitPlotCorrect1}
\end{figure}

The first direct check is to compare the dispersion integral \eqref{M1EDispequation} with numerical data of $\mathcal M_1(\eta)$ at real $\eta$. Approximated $\Delta_1(y)$ from interpolation (see Fig.\ref{Continued_Disc_Delta1}) are used for the numerical integration in \eqref{M1EDispequation}, and \eqref{M1EDispequation} is to be compared with numerical data of $\mathcal M_1(\eta)$ measured from the truncated spectrum on the interval $ -4.0 < \eta < + 4.0$. The data of $\mathcal M_1(\eta)$ are with 4 to 5 digits accuracy. The results of comparison are as shown in Fig.\ref{M1EdispPlotCorrect1} and Fig.\ref{M1EdispMunitPlotCorrect1}, with presenting scaling functions $\mathcal M_1$ and $\hat M_1$ ($M_1$ measured in unit of $|h|^{8/15}$ or $|m|$) respectively. Both Fig.\ref{M1EdispPlotCorrect1} and Fig.\ref{M1EdispMunitPlotCorrect1} are showing agreement in good accuracy on the interval of $-4.0 \le \eta \le +4.0$. The evidence is supporting the conjecture of extended analyticity.

Apart from direct comparing the $M_1$ data with the integral \eqref{M1EDispequation}, another independent verification can be made by checking expansion coefficients at small $\eta$. By expanding the integration \eqref{M1EDispequation}, coefficients $M_1^{(n)}$ (see \eqref{Mp_Eta_expansion1}) or \eqref{Mp_expansion_mn} can be represented as dispersion integrations. On the other hand, some of $M_1^{(n)}$ are known and can be served as consistency checks. By using the discontinuities shown in Fig.\ref{Continued_Disc_Delta1}, these integrations read:
\begin{gather}
M_1^{(2)} = 0.2002 \quad \text{while} \quad -\frac{2}{\pi}\int_{Y_0}^{+\infty} \frac{dy}{y^3} \Re e \, \big( e^{\frac{4 \pi i}{15}} \Delta_1(y) \big) \approx 0.2051 \,, \label{M12DispCoeff}\\
M_1^{(3)} = -0.051 \quad \text{while} \quad +\frac{2}{\pi}\int_{Y_0}^{+\infty} \frac{dy}{y^4} \Re e \, \big( e^{\frac{8 \pi i}{15}} \Delta_1(y) \big)  \approx -0.050 \,, \label{M13DispCoeff}\\
M_1^{(4)} \,\, \text{is unknown,} \quad \text{while} \quad +\frac{2}{\pi}\int_{Y_0}^{+\infty} \frac{dy}{y^5} \Re e \, \big( e^{-\frac{\pi i}{5}} \Delta_1(y) \big)  \approx -0.0083 \,,\label{M14DispCoeff}
\end{gather}
where $M_1^{(2)} $ was given in Tab.\ref{Tab:Mp_coefficients}, $M_1^{(3)}$ was also measured in \cite{Xu:2022mmw}, and there's no known numerical value of $M_1^{(4)}$ available. The above comparison of $M_1^{(2)}$ and $M_1^{(3)}$ are showing good agreement, with deviations less than $3\%$.

\begin{figure}[!th]
\centering
\includegraphics[width=0.8\textwidth]{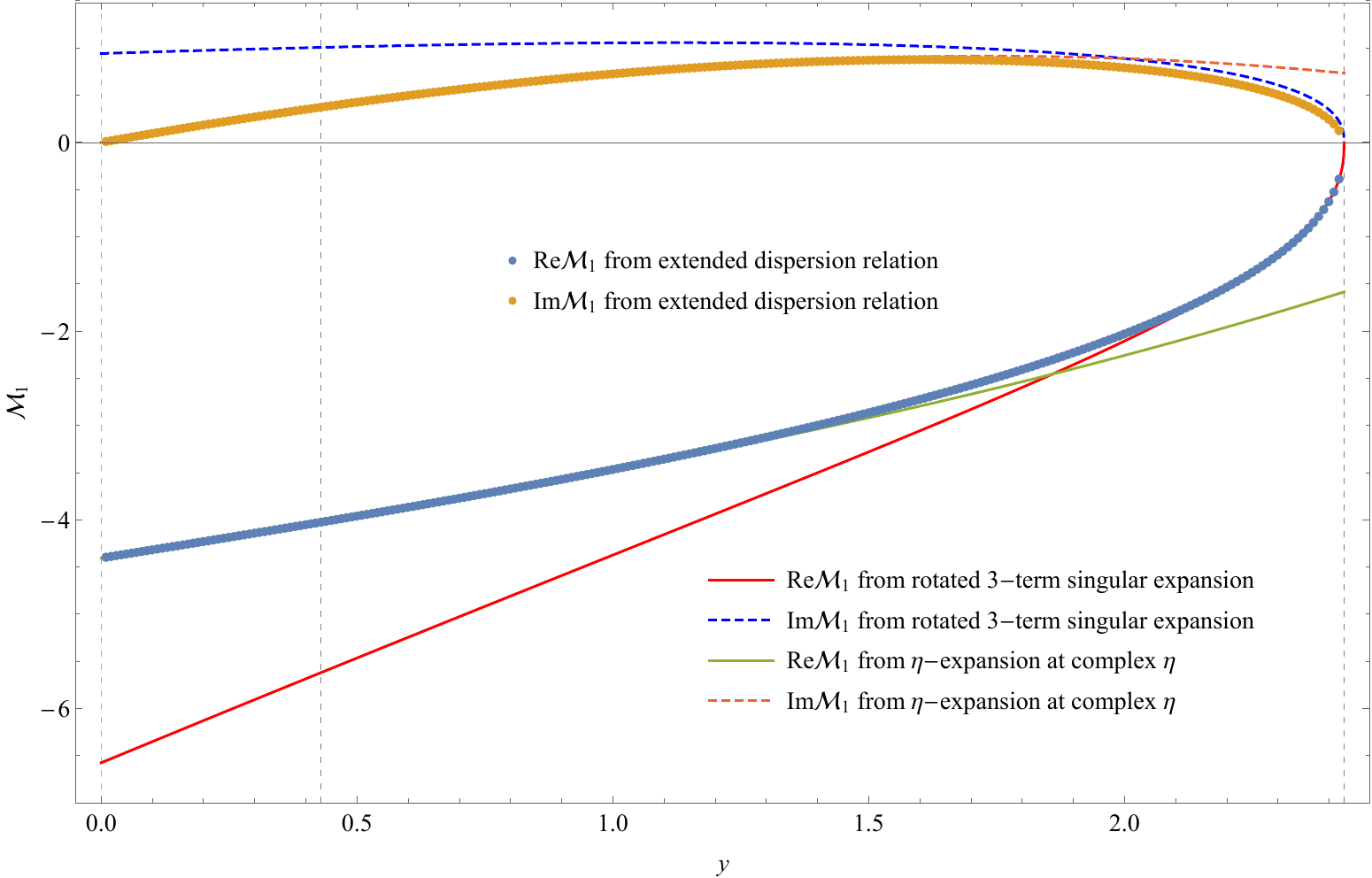}
\caption{A verification of $M_1$ extended dispersion relation at complex $\eta = ye^{\frac{11}{15}\pi i}$. Dispersion integral \eqref{M1EDispequation} is plotted for $\eta = ye^{\frac{11}{15}\pi i}$ and $ 0 < y <Y_0$, with real and imaginary parts as bullets in different colors. They are to be compared to the series expansions at $y=Y_0$ and $y=0$. Near the Yang-Lee point, the real and imaginary parts of \eqref{M1SingularRotated} are plotted as red solid curve and blue dashed curve. Near $y = 0$, \eqref{M1regularExpansion1} is plotted with $\eta \to y e^{\pm \frac{11 \pi i}{15}}$, with the truncation till the term with $M_1^{(3)}$. The real and imaginary part are plotted as green solid curve and pink dashed curve. The dispersion integral \eqref{M1EDispequation} matches well with \eqref{M1SingularRotated} near $y=Y_0$, and with \eqref{M1regularExpansion1} near $y=0$.}
\label{M1EdispImaginaryH1}
\end{figure}

Furthermore, we should check \eqref{M1EDispequation} not only at real $\eta$ but also works when $\eta$ is complex. The verification can be made by computing \eqref{M1EDispequation} for $\mathcal M_1(\eta)$ at $\eta = - y e^{\pm\frac{4}{15}\pi i}$ with $0 < y < Y_0$. On this interval between the Yang-Lee point and the $E_8$ point, the function $\mathcal M_1(\eta = -ye^{\mp\frac{4\pi i}{15}})$ is complex, and can be related to the values of $\hat M_1(\xi^2)$ on the upper/lower edges of the Yang-Lee branch cut, as \eqref{Mcal1WrongSideDef}. Unfortunately, the measurement of $\hat M_1(\xi^2 \pm i0)$ for $\xi^2 < -\xi^2_0$ are with poor accuracy (see discussions and Fig.18 in \cite{Xu:2022mmw}), thus we shall verify complex \eqref{M1EDispequation} by comparing it to the expansions at $\eta = 0$ and $\eta = -Y_0 e^{\pm\frac{4}{15}\pi i}$ instead, with both real and imaginary parts. Near the Yang-Lee point the numerical integration is to be compared with:
\[
\bar{\mathcal M}_1(y) = (Y_0 - y)^{\frac{5}{12}}\big[ \tilde{b_0} e^{+\frac{5\pi i}{12}} + \tilde{c_0} (Y_0 - y)^{\frac{5}{6}} e^{+\frac{5\pi i}{4}} + \tilde{b_1}  (Y_0 - y) e^{+\frac{17\pi i}{12}} + \cdots \big] \,. \label{M1SingularRotated}
\]
based on \eqref{M1Singular}, and near $y = 0$ the $E_8$ point the numerical integration is be compared with \eqref{M1regularExpansion1} with complex $\eta = y e^{\pm \frac{11 \pi i}{15}}$. The results are as shown in Fig.\ref{M1EdispImaginaryH1}, with good accordance on the interval $0\le y \le Y_0$ for both real and imaginary part.

In summary, the verifications of $M_1$ extended dispersion relation are given at real $\eta$, as comparing the truncated data of $\mathcal M_1(\eta)$ to the numerical integration \eqref{M1EDispequation}, see Fig.\ref{M1EdispPlotCorrect1} and Fig.\ref{M1EdispMunitPlotCorrect1}). These results, combined with two other cross checks, as comparing the expansion coefficients $M_1^{(k)}$ (see \eqref{M12DispCoeff}, \eqref{M13DispCoeff} \& \eqref{M14DispCoeff}), and as comparing \eqref{M1EDispequation} with complex $\eta$ to the expansions \eqref{M1Singular} and \eqref{M1regularExpansion1} (see Fig.\ref{M1EdispImaginaryH1}). All this evidence strongly support the extended analyticity conjecture of $M_1$, as illustrated in Fig.\ref{eta_planes}(b), that no other singularity exist within the shadowed domain or along the rotated Yang-Lee branch cut.

\section{Summary and discussion}
In this work, we continue the study on IFT spectrum of the first mass $M_1$ (see \cite{Xu:2022mmw}). Analyticity of $M_1$ was conjectured in the low-T regime, as in Fig.\ref{LangerBC} and \eqref{M123_LowT_disperion_eq}. Numerical approximation of the discontinuities along the Fisher-Langer's branch cut was established (see Fig.\ref{M123_LowT_discontinuities_piecewise1}), and with its help, the low-T analyticity was verified from various aspects, as in Fig.\ref{M123_LowT_disp_integrals_checks},
and \eqref{M123_LowT_disp_integrals_checks}. Furthermore, on the complex $\eta$-plane the extended analyticity conjecture of $M_1$ was formulated, which unifies the behaviours of $M_1$ in both high-T and low-T regimes. The complex structure of $\mathcal M_1(\eta)$ on the complex $\eta$ plane is as shown in Fig.\ref{eta_planes}(b), with the rotation of Yang-Lee branch cut. The associated extended dispersion relation of $M_1$ was given in \eqref{M1EDispequation}, which depends on the discontinuities $\Delta_1(y)$ along the rotated Yang-Lee branch cut. The approximation of $\Delta_1(y)$, based on the expansions \eqref{Delta1SingularExpansion1} and \eqref{Delta1AsymptoticExpansion1}, are presented in Fig.\ref{Continued_Disc_Delta1}, with the details given in Appendix \ref{Appendix_B}. With its help, the extended dispersion relation \eqref{M1EDispequation} were verified with various comparisons, as with the truncated data (Fig.\ref{M1EdispPlotCorrect1} \& Fig.\ref{M1EdispMunitPlotCorrect1}), of the $\eta$-expansion coefficients (\eqref{M12DispCoeff} \& \eqref{M13DispCoeff}), and at complex $\eta$ to the expansions \eqref{M1SingularRotated} \& \eqref{M1regularExpansion1} (see Fig.\ref{M1EdispImaginaryH1}). All of which show good accordance (deviations $ < 3\%$), and support the conjecture of $M_1$ extended analyticity conjecture.

Our extended analyticity conjecture of $M_1$ is in agreement with the one of the free energy density, which was conjectured and verified in \cite{fonseca2003ising}. Both extended analyticity conjectures state that no other singularity exist within the shadowed domain (domain of $ \frac{8}{15} \pi < |\text{Arg}\, \eta| < \frac{11}{15}\pi$, see Fig.\ref{eta_planes}(b)). However, there are several major differences between the $M_1$ extended dispersion relation \eqref{M1EDispequation} and the dispersion relation of free energy density (see (4.8) of \cite{fonseca2003ising}). Firstly, the order of divergences at $|\eta| \to \infty$ and subtractions are different. $\mathcal M_1(\eta)$ is divergence with $O(\eta)$ at $|\eta| \to \infty$, thus the subtractions are of $M_1^{(0)} + M_1^{(1)} \eta$. Meanwhile, function $\tilde \Phi(\eta)$ (the scaling function of free energy density, defined in (3.20) \& (3.25) of \cite{fonseca2003ising}) has more complicated behaviours at $|\eta| \to \infty$, as a combination of Onsager's singularity $\sim \eta^2 \log \eta^2$ and series expansion in fractional powers of $\eta$ (see (3.35) \& (3.36) of \cite{fonseca2003ising}). As a result, the extended dispersion relation of free energy density is much more complicated comparing to \eqref{M1EDispequation}. Secondly, the discontinuities along the rotated Yang-Lee branch cut have very different behaviours. According to the analysis in Sec.\ref{Section3} and Appendix \ref{Appendix_A}, the contributions to $\Delta_1(y)$ at $y \to +\infty$ are mostly controlled by the expansions of $\hat M_1$ at small $\xi$, with the instanton-like contribution negligible. However, the discontinuity of $\tilde \Phi(\eta)$ is significantly contributed by the instanton-like terms ($\sim y^{1/8} \exp\big( - \frac{\pi}{2\bar s} e^{-\frac{3\pi i}{8}} y^{15/8}\big)$, see (8.19) of \cite{fonseca2003ising}). The reason was explained in the introduction between \eqref{FLowExpansion} and \eqref{FmetaCondensation}, as since \eqref{FLowExpansion} has vanishing continuation along the Fisher-Langer's branch cut, the discontinuity of free energy density is determined by the instanton-like term \eqref{FmetaCondensation}, and same as it continued across the shadowed domain.

In Sec.\ref{Section2}, we also analyzed the analyticity of functions $\hat M_2(\xi)$ and $\hat M_3(\xi)$ in the low-T $\xi$-plane, which are as shown in Fig.\ref{LangerBC}. However, for the second mass $M_2$ and third mass $M_3$, the analyticities of their corresponding scaling functions $\mathcal M_2(\eta)$ and $\mathcal M_3(\eta)$ on the complex $\eta$ plane, if exist, should be more involved, comparing to the $M_1$ extended analyticity conjecture and the dispersion relation \eqref{M1EDispequation}. The reason is that in the high-T regime $T > T_c$, at finite $\xi_2$ and $\xi_3$, the masses of second particle and third particle exceed $2M_1$, and the particles become unstable. For $\xi^2 < \xi^2_2$ or $\xi^2 < \xi^2_3$, the scaling functions $\hat M_2(\xi^2)$ and $\hat M_3(\xi^2)$ are neither measurable via truncated method nor have good analyticity properties. Furthermore, little is known about what the Yang-Lee singularities means for the higher particles, not to mention the behaviours of discontinuities of associated scaling functions (if not using $\hat M_p$ or $\mathcal M_p$) along the Yang-Lee branch cut. Still, there exist evidences of a continuous evolution of the Ising mass spectrum along the real $\eta$ axis \cite{zamolodchikov2013ising}, which suggest the possibility of higher masses extended analyticities with much more involved forms. We hope to return to this topic in the future.

\section*{Acknowledgments}
The author is most thankful to A. B. Zamolodchikov for illuminating discussions and advices on the manuscript. I also thank P. Vieira, R. Shrock and B. McCoy for helpful discussions. Research of HLX is partly supported by NSF under grant PHY-2210533.


\appendix
\section*{Appendix}

\section{Numerical approximation of $\Delta_1(y)$ from interpolation}\label{Appendix_B}
In this section, we will provide details of the interpolation which gives the numerical approximation of $\Re e \, \Delta_1(y)$ and $\Im m \, \Delta_1(y)$, as shown in Fig.\ref{Continued_Disc_Delta1}. The interpolation method is similar to what were discussed in \cite{fonseca2003ising},\cite{Xu:2022mmw} and\cite{Xu:2023nke}.

Following \eqref{Delta1AsymptoticExpansion1} and using parameter $\chi = 1/y$, $\mathcal E_1(\chi) = \chi \Delta_1(y = \frac{1}{\chi})$ admits the asymptotic expansion at $\chi = 0$, as:
\begin{gather}
\mathcal E_1(\chi) = \frac{3}{2}i + \frac{i}{2} \sum_{k=1}^\infty a_k \chi^{\frac{5}{4} k } e^{-\frac{11}{12}k\pi i}
- \frac{i}{2} \sum_{k=1}^\infty \mu_{2k} \chi^{\frac{15}{4} k } e^{-\frac{11}{4}k\pi i}  \,. \label{EcalExpansion}
\end{gather}
Define $t = \chi^{\frac{5}{4}}$, \eqref{EcalExpansion} becomes:
\begin{gather}
\Im m \, \mathcal E_1 = \frac{3}{2} + \frac{1}{2} \sum_{k=1}^\infty a_k \,t^k  \cos \Big( \frac{11 \pi k}{12} \Big)- \frac{1}{2} \sum_{l=1}^\infty \mu_{2l}\,  t^{3l} \cos \Big( \frac{11 \pi l}{4} \Big) \,, \label{E1tImExpansion}\\
\Re e \, \mathcal E_1 =  \frac{1}{2} \sum_{k=1}^\infty a_k\, t^k \sin \Big( \frac{11 \pi k}{12} \Big) - \frac{1}{2} \sum_{l=1}^\infty \mu_{2l}\,  t^{3l} \sin \Big( \frac{11 \pi l}{4} \Big) \,,\label{E1tReExpansion}
\end{gather}
which works near $t = 0$. On the other hand, near the Yang-Lee point $y=Y_0$ or equivalently $t = T_0 := Y_0^{-\frac{5}{4}}$, $\mathcal E_1$ admits singular expansion in powers of $(T_0 - t)$:
\begin{gather}
\mathcal E_1 = {(T_0 - t)^{\frac{5}{12}}} \Big\{  \hat{b_0} \sin(\frac{5\pi}{12}) e^{\frac{5\pi i}{12}} + \hat{c_0} \sin(\frac{5\pi}{4}) e^{\frac{5\pi i}{4}}(T_0 - t)^{\frac{5}{6}}  \nonumber\\
  + \hat{b_1} \sin(\frac{17\pi}{12}) e^{\frac{17\pi i}{12}} (T_0 - t) + \cdots \Big\}\,, \label{E1tSingular}
\end{gather}
which follows \eqref{Delta1SingularExpansion1}, with the coefficients $\hat b_0$, $\hat b_1$ and $\hat c_0$ were computed from \eqref{b0b1c0values1}, as:
\[
\hat b_0 = 2.6511 \,, \quad
\hat b_1 = -1.121 \,, \quad
\hat c_0 = -3.558 \,. \label{b0b1c0Hatvalues1}
\]
\eqref{E1tSingular} can be further extended with more terms, and the exponents of the additional terms follow the operator contents of Yang-Lee CFT and dimensional analysis \cite{Xu:2022mmw}\cite{Xu:2023nke}. In addition to \eqref{E1tSingular}, we add 4 more terms as:
\begin{gather}
\mathcal E_1^{\text{Approx}} = \eqref{E1tSingular} + \hat{e_0} \sin(\frac{25\pi}{12}) e^{\frac{25\pi i}{12}}(T_0 - t)^{\frac{25}{12}}
+ \hat{c_1} \sin(\frac{9\pi}{4}) e^{\frac{9\pi i}{4}}(T_0 - t)^{\frac{9}{4}} \nonumber\\
+ \hat{b_2} \sin(\frac{29\pi}{12}) e^{\frac{29\pi i}{12}}(T_0 - t)^{\frac{29}{12}}
+ \hat{d_0} \sin(\frac{11\pi}{4}) e^{\frac{11\pi i}{4}}(T_0 - t)^{\frac{11}{4}} \,,
 \label{E1tSingularExtended}
\end{gather}
with $\hat e_0$, $\hat c_1$, $\hat b_2$ and $\hat d_0$ are four unknown coefficients. These coefficients are determined by requiring 4 constraints as:
\[
\mathcal E_1^{\text{Approx}} \Big|_{t= 0 } = \mathcal E_1 \Big|_{t= 0 } \,, \quad \text{and} \quad
\frac{d}{dt}\mathcal E_1^{\text{Approx}} \Big|_{t= 0 } = \frac{d}{dt}\mathcal E_1 \Big|_{t= 0 } \,,
\]
for real and imaginary parts individually. The right hand sides follow the expansions \eqref{E1tImExpansion} and \eqref{E1tReExpansion} at $t=0$, with:
\begin{gather*}
\Re e \,\mathcal E_1 \Big|_{t=0}= 0 \,, \quad
\frac{d}{dt}\Re e \,\mathcal E_1 \Big|_{t=0}= -0.5844 \,, \\
\Im m \,\mathcal E_1 \Big|_{t=0}= 1.5 \,, \quad
\frac{d}{dt}\Im m \,\mathcal E_1 \Big|_{t=0}= -2.181 \,,
\end{gather*}
the numerical solution of extra coefficients are:
\[
\hat e_0 = -24.69\,, \quad
\hat c_1 = -15.83\,, \quad
\hat b_2 = 27.38\,, \quad
\hat d_0 = -32.30\,. \label{ExtraSingularCoefficientsValues}
\]
As a result, we take the 7-term expansion \eqref{E1tSingularExtended}, where the coefficients are \eqref{b0b1c0Hatvalues1} \& \eqref{ExtraSingularCoefficientsValues}, with the parameter $t$ transformed back to $\chi = t^{4/5} = 1/y$, as the numerical approximations of $\mathcal E_1 (\chi)$ along the rotated Yang-Lee branch cuts. Real and imaginary parts of the approximation $\mathcal E_1^{\text{Approx}} (\chi)$ were the ones as shown in Fig.\ref{Continued_Disc_Delta1}, and reproduce the numerical verification of $M_1$ extended dispersion relation as was discussed in Sec.\ref{Section3}.

\section{Computing the non-analytic term $\mathcal K_p$.}\label{Appendix_A}
In this section, we shall analyze the "instanton-like" contribution of $M_1$ in the low-T regime and its continuation. The analysis follows the theory of condensation of nucleation, see \cite{fisher1967theory,langer2000theory,andreev1964singularity,gunther1980goldstone,lowe1980instantons,harris1984ising,Voloshin:1985id} for more detailed discussions.

As was discussed in the introduction, when $T < T_c$ and exist a small nonvanishing external magnetic field $h$, the double degenerate vacua \eqref{sigmaVEV} are shifted, splitted into the stable vacuum (spins align along the external field) and metastable vacuum (spins align against the external field). The continuation $h \to e^{\pm\pi i} h$ can be interpreted as flipping the direction of external magnetic field, thus the roles of stable and metastable vacua are interchanged. As a result, thermodynamic quantities at $\xi \to 0^-$ should be understood as of those in the metastable phase. For example, $M_1(m, -h \pm i0)$ should be understood as the inverse correlation length in the metastable phase, and is contributed by both the term-by-term continuation the term-by-term continuation \eqref{MesonSeriesExpansion2} and the instanton-like term. Here we denote the term-by-term contribution of $M_1(m, -h \pm i0)$ as $\tilde{M}^{(\pm)}_1 = \tilde{M}^{(\pm)}_1(m,\mathfrak h) =  M_1(m,e^{\pm\pi i} \mathfrak h)$, where $\mathfrak h$ is the external field\footnote{On the upper/lower edge of Fisher-Langer's branch, $h = e^{\pm\pi i} \mathfrak{h}$ with $\mathfrak{h} > 0$.}, and the instanton-like term $K_1(m,\mathfrak h)$ is defined as:
\[
{M}_1(m , - h \pm i0) =  \tilde{M}^{(\pm)}_1(m,\mathfrak h)  \pm  K_1(m,\mathfrak h)\,. \label{Instanton-like_term_def}
\]
The dimensionless scaling function of $K_1$ is defined by:
\[
\mathcal K_1(\tilde \lambda) = \frac{1}{m} K_1(m,\mathfrak h) \,, \quad \text{where} \quad \label{Kappa_term_def}
\tilde \lambda = \frac{2\bar \sigma \mathfrak h}{m^2} = 2 \bar s x > 0\,,
\]
and $x = \mathfrak h / m^{15/8} = -\xi > 0$.

At finite temperature, the metastable vacuum is perturbed by thermal fluctuations. Microscopically, thermal fluctuations generate clusters of spins align along the external field, with satisfying certain distributions. Among them, sufficiently large clusters would stabilize their shapes as round circles, and can be interpreted as bubbles of stable vacuum. Assume the bubbles are circular with radius $R$, the one bubble effective action reads \cite{Voloshin:1985id}:
\[
{\mathcal A}_0  = +2 \pi m R - 2 \pi \bar{\sigma} \mathfrak{h} R^2 = +\pi m R_c - 2 \pi \bar{\sigma} \mathfrak{h} (R-R_c)^2 \,, \label{BubbleActionSB}
\]
which is contributed by the area term (due to the surface tension) and volume term (due to the overall energy density difference between the vacua). Here the critical radius is defined as
\[
R_c = \frac{m}{2\bar{\sigma} \mathfrak{h}} = \frac{1}{ m \tilde\lambda} = \frac{1}{2\bar s x \cdot m}\,,
\]
which is the saddle point of \eqref{BubbleActionSB}. For later convenience, \eqref{BubbleActionSB} can be written in dimensionless form, with $r = m R$:
\[
{\mathcal A}_0 = +\frac{\pi}{\tilde\lambda} - {\pi}{\tilde\lambda} \big( r - \frac{1}{\tilde\lambda} \big)^2 \,.
\]

However, with the negative quadratic term of \eqref{BubbleActionSB}, any bubble with size $R > R_c$ would explode instantly due to instability, and lead to the decay of metastable vacuum. Such process is known as nucleation, and provides the physical interpretation of \eqref{FmetaCondensation} as the metastable vacuum decay rate. The instability nature of metastable vacuum suggests that computations regarding instanton-like terms are only defined in terms of analytical continuation, and further subtleties will be discussed in this section.

In fact, the effective action \eqref{BubbleActionSB} is only a rough description, because it ignores the thickness of bubble shell and the variations of bubble shape. The thickness of bubble shell can be interpreted as the size of instantons, which tunnel between the vacuums, and when $\tilde\lambda$ is small, or to say $2 \bar\sigma \mathfrak h \ll m^2$, the tunneling size is negligible. Analysis of the latter one requires path-integration over all possible bubble shapes \cite{Voloshin:1985id}, which is as follows.

To include the fluctuations of bubble shape, angular dependence is added to the radius, as using the polar coordinate $(\mathcal R, \alpha)$\footnote{In \cite{Voloshin:1985id}, it was argued that at leading order $\mathcal R(\alpha)$ is a single-valued function of $\alpha$. The extremely concave bubble gives higher order singularity. For example, dumbbell-like bubble contributes terms $\sim e^{-{\pi}/{\tilde\lambda^3}}$.}. The one bubble effective action \eqref{BubbleActionSB} becomes:
\[
\mathcal A[\mathcal R] = m \int_{0}^{2\pi} \sqrt{\dot{\mathcal R}^2 +{\mathcal R}^2} d\alpha - 2 \bar{\sigma} \mathfrak h \int_{0}^{2\pi} \frac{1}{2} {\mathcal R}^2  d\alpha \,, \label{BubbleAction1}
\]
where we are replacing $ R \to \mathcal R = \mathcal R(\alpha)$ and abbreviating $\dot{\mathcal R} =\frac{d}{d\alpha} \mathcal R(\alpha)$. The radius coordinate $\mathcal R$ can be decomposed near the stationary configuration, as $\mathcal R(\alpha) = R_c + \rho(\alpha)$, where $\rho(\alpha)$ represents small deviation from $R_c$.

To analyze the effect of bubbles on $M_1$, we consider the 2-point correlation function in metastable vacuum, which is defined as:
\[
G(L) = \langle \sigma(L) \sigma(0) \rangle = \frac{\int \mathcal{D}\sigma e^{-\mathcal A} \sigma(L) \sigma(0) }{\int \mathcal{D}\sigma e^{-\mathcal A}} \,, \label{2ptCorrelatorDef}
\]
where all possible $\sigma(x)$ configurations have been path-integrated
. In the metastable vacuum, the path integrals of \eqref{2ptCorrelatorDef} can be expanded as summation over clusters of spins align along to the external field, equivalent to the bubbles of \eqref{BubbleAction1}. With shorthand notation, we denote the denominator of \eqref{2ptCorrelatorDef} as:
\begin{gather}
\mathcal N = \int \mathcal{D}\sigma e^{-\mathcal A} = \sum_{n=0}^{\infty}\mathcal N_n =  \mathcal N_0 + \sum_{n=1}^{\infty}\frac{1}{n!} \prod_{k=1}^{n} \Big( \int_{-\infty}^{+\infty} dx_k dy_k \int_{\gamma_{k}} \mathcal D{\mathcal R}_k \Big) e^{-\sum_{k}\mathcal A[\mathcal R_k]} \,,\label{NdenominatorExpansion}\\
\mathcal N_n = \frac{1}{n!} \Big( \int_{-\infty}^{+\infty} dx dy \int_{\alpha = 0}^{\alpha = 2\pi} \mathcal D{\rho}(\alpha) e^{-\mathcal A[\mathcal R]}\Big)^n\,, \label{NndenominatorDef}
\end{gather}
where in \eqref{NdenominatorExpansion} we have expanded $\mathcal N$ by the number of bubbles, and for each bubble its coordinate $(x_k,y_k)$ integrated. The simple closed curve $\gamma_k$ is to describe the shape of bubble with labelling $k$, and in \eqref{NndenominatorDef} the integration over bubble configurations $\int \mathcal{D} \mathcal R$ is replaced by $\int \mathcal{D} \rho$. Same as $\mathcal R(\alpha)$, function $\rho(\alpha)$ is also single-valued and periodic, with $\rho(0) = \rho(2\pi)$. Similarly, expanding the numerator of \eqref{2ptCorrelatorDef} gives:
\begin{gather}
\mathcal G (L) = \int \mathcal{D}\sigma e^{-\mathcal A}  \sigma(L) \sigma(0) =  \mathcal G_0(L) + \sum_{n=1}^{\infty} \mathcal G_n(L) \,,\label{GdenominatorExpansion}\\
\mathcal G_n(L) = \frac{1}{n!} \prod_{k=1}^{n} \Big( \int_{-\infty}^{+\infty} dx_k dy_k \int_{\alpha = 0}^{\alpha = 2\pi} \mathcal D{\rho}_k(\alpha)  \Big) e^{-\sum_k\mathcal A[\mathcal R_k]} \sigma(L) \sigma(0) \,. \label{GndenominatorDef}
\end{gather}
Obviously $\mathcal G(L)$ and each $\mathcal G_n(L)$ are only functions depending on $L$, and $\mathcal G_0$ should proportional to the spacetime volume $\text{Vol}(\mathbb R^2)$.

\begin{figure}[!t]
\centering
\includegraphics[width=0.8\textwidth]{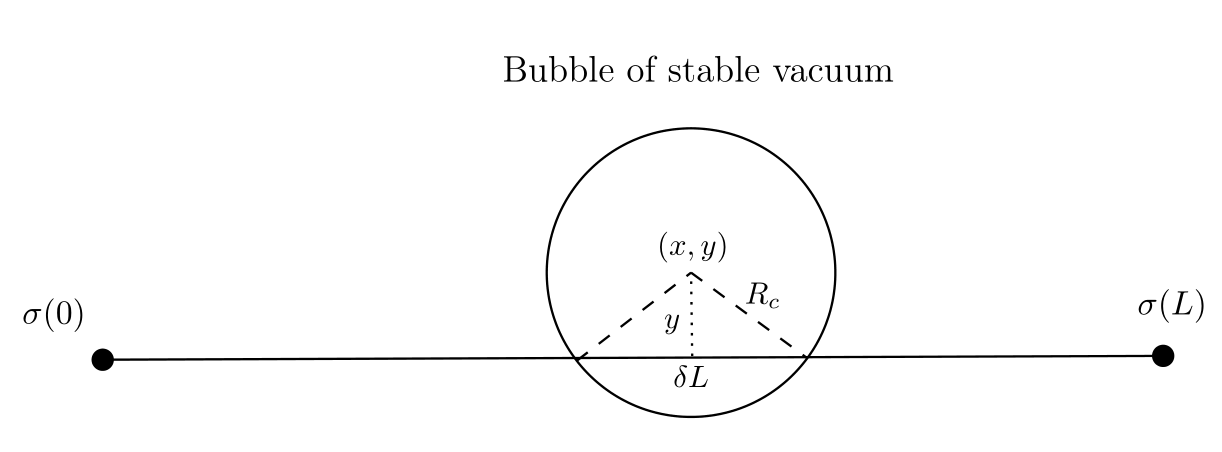}
\caption{The contribution of one critical bubble to $\mathcal G_1(L)$ in \eqref{1BubbleCorrelator0}. The two spin operator are inserted at $(0 , 0)$ and $(L, 0 )$, and the center of critical bubble is located at $(x,y)$. The path of particle propagating within the bubble is with length $\delta L = 2 \sqrt{R_c^2 - y^2}$}
\label{DropletCorrectionPicture}
\end{figure}

Based on \eqref{NndenominatorDef} and \eqref{GndenominatorDef}, the 2-point correlator of the metastable phase \eqref{2ptCorrelatorDef} can be rearranged by the number of bubbles, as:
\begin{gather}
\langle \sigma(L) \sigma(0) \rangle = \frac{\mathcal G_0}{\mathcal N_0} + \frac{\mathcal G_1 -  \frac{\mathcal N_1}{\mathcal N_0}\mathcal G_0}{\mathcal N_0} +
\frac{\mathcal G_2 - \big(\frac{ \mathcal N_2}{\mathcal N_0} - \frac{\mathcal N_1^2}{\mathcal N_0^2} \big)\mathcal G_0 - \frac{ \mathcal N_1}{\mathcal N_0} \mathcal G_1 }{\mathcal N_0} + \cdots  \\
= G^{(0)}(L) + G^{(1)}(L) + G^{(2)}(L) + \cdots \,, \label{2ptCorrelatorBubbleExpansion}
\end{gather}
with $G^{(k)}(L)$ represents the normalized $k$-bubble contribution. The leading term $G^{(0)} = \mathcal G_0 / \mathcal N_0$ is finite, as both $\mathcal G_0$ and $\mathcal N_0$ are with divergences proportional to $\text{Vol}(\mathbb R^2)$. $G^{(1)}(L)$ represents the one bubble correction, with both $\mathcal G_1$ and $\mathcal N_1$ diverge with $\big(\text{Vol}(\mathbb R^2)\big)^2$, due to the extra integral over the position of bubble. However $G^{(1)}(L)$ remains finite, by the cancellation of divergences between the two terms of its numerator.

\begin{figure}[!htp]
\centering
\includegraphics[width=0.8\textwidth]{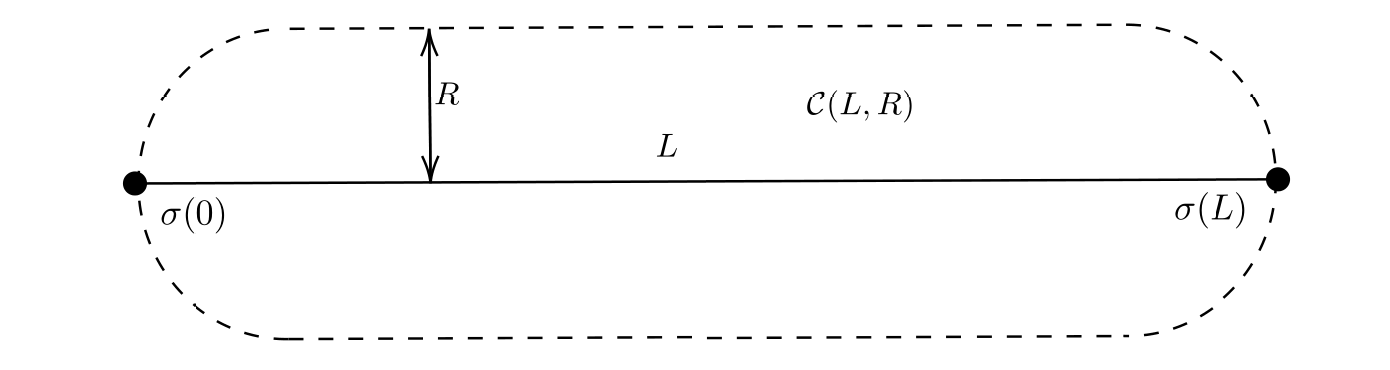}
\caption{The cigar-like integration domain $\mathcal C(L,R)$ in \eqref{1BubbleCorrelator1}.}
\label{CigarPicture}
\end{figure}

To compute $G^{(1)}(L)$, consider the one bubble configuration shown as in Fig.\ref{DropletCorrectionPicture}\footnote{In the low-T regime, the lightest particle is the first meson of the McCoy-Wu scenario. The size of first meson can be understood as the maximum separation between the semiclassical trajectories of quarks, which is $X_1 = 2 R_c (\cosh \vartheta_1(\lambda) -1)$. At small $\lambda$, $X_1 \propto 1/\lambda^{1/3}$. The size of meson is much smaller than the size of critical bubble, as $X_1 \ll R_c$. As a result, in Fig.\ref{DropletCorrectionPicture} the size of propagating particle is negligible.}. The bubble is perturbed near its critical radius, with $\mathcal R  = R_c +\rho$, and we denoted the center of the bubble as $(x,y)$. In Fig.\ref{DropletCorrectionPicture}, the path propagating within the stable vacuum bubble is with length $\delta L$, and the one bubble contribution reads:
\begin{gather}
G^{(1)}(L) = \frac{1}{\mathcal N_0} \Big( \mathcal G_1 (L) -  \frac{\mathcal N_1}{\mathcal N_0}\mathcal G_0(L)  \Big) \label{1BubbleCorrelator0}\\
= \frac{1}{\text{Vol}(\mathbb R^2)} \int_{\mathcal C(L,R)} dx dy \int \mathcal D \rho(\alpha)e^{-\mathcal A(\mathcal R)} \cdot
\frac{e^{-\tilde{M}_1 L}}{L^{1/2}}\Big( e^{(\tilde{M}_1 - {M}_1)\delta L} - 1 \Big)\,, \label{1BubbleCorrelator1}
\end{gather}
where the factor $ e^{(\tilde{M}_1 - {M}_1)\delta L}$ is the result of the particle propagating within the bubble\footnote{The superscript sign $(\pm)$ of $\tilde M_1$ representing upper/lower edge is suppressed here.}, and the $-1$ term in \eqref{1BubbleCorrelator1} follows the second term $-  \frac{\mathcal N_1}{\mathcal N_0^2}\mathcal G_0(L)$ of \eqref{1BubbleCorrelator0}. The integration regime of $\int dxdy$ is the cigar-like domain $\mathcal C(L,R_c)$ as shown in Fig.\ref{CigarPicture}. When the center of bubble lies outside of the cigar, as $(x,y) \notin \mathcal C(L,R_c)$, two terms in \eqref{1BubbleCorrelator0} cancel with each other. When $L \gg R_c$, the cigar can be approximated as a rectangular, and \eqref{1BubbleCorrelator1} becomes:
\[
G^{(1)}(L) = \frac{1}{\text{Vol}(\mathbb R^2)} \int_{0}^{L}dx \int_{-R_c}^{+R_c} dy \int \mathcal D \rho \, e^{-\mathcal A[\mathcal R]}
\frac{e^{-\tilde{M}_1 L}}{L^{1/2}}\Big( e^{(\tilde{M}_1 - {M}_1)\delta} - 1 \Big)\,, \label{1BubbleCorrelator2}
\]
with $\delta = \delta(y,R_c) = 2 \sqrt{R_c^2 - y^2}$.

\eqref{1BubbleCorrelator2} has dealt with the one bubble contribution, and if ignoring interactions between the bubbles, the multi-bubble contribution can be treated with similar manner, with the summation of \eqref{2ptCorrelatorBubbleExpansion} exponentiates. As a result, \eqref{1BubbleCorrelator2} is related to the instanton-like term $\mathcal K_1$ (defined in \eqref{Kappa_term_def}) by (on the upper edge):
\[
G^{(1)}(L) = - (mL) \frac{e^{-\tilde M_1 L}}{\sqrt{L}} \cdot \mathcal K_1(\lambda) \,. \label{Kappa_from_G1}
\]
and we will compute $\mathcal K_1(\lambda)$ by doing appropriate regularization of \eqref{1BubbleCorrelator2}.

Back to \eqref{1BubbleCorrelator2}, the integration of $dxdy$ over the rectangle becomes:
\[
\int_0^L d x \int_{-R_c}^{+R_c} d y \Big( e^{N_1\sqrt{R_c^2 - y^2}} - 1 \Big)= \pi L R_c \Big( I_1( N_1 R_c ) + L_{-1}( N_1 R_c ) - \frac{2}{\pi} \Big) \,,
\]
where $N_1 = N_1 (m, \mathfrak h) = 2\tilde{M}_1 - 2M_1$, and $I_1(z)$ \& $L_{-1}(z)$ are the Bessel-I function and Struve-L function. \eqref{1BubbleCorrelator2} then becomes a path integration over $\mathcal D \rho$, namely the variations of the bubble shape, as:
\[
G^{(1)}(L) = \frac{\pi L}{\text{Vol}(\mathbb R^2)} \frac{e^{-\tilde{M}_1 L}}{L^{1/2}} \int \mathcal D \rho \, e^{-\mathcal A[\mathcal R]}R_c
\Big(I_1( N_1 R ) + L_{-1}( N_1 R ) - \frac{2}{\pi}\Big)\,, \label{1BubbleCorrelator3}
\]
and similar to the computation regarding free energy in \cite{langer2000theory}\cite{Voloshin:1985id}, the integral $\mathcal D\rho$ in \eqref{1BubbleCorrelator3} counts the partition sums in metastable vacuum. In fact, \eqref{1BubbleCorrelator3} is a gaussian-like path integration, because the action $\mathcal A[\mathcal R]$ can be reduced to the action of compact boson. Expanding \eqref{BubbleAction1} near the stationary radius $\mathcal R(\alpha) = R_c + \rho(\alpha)$:
\begin{gather}
\mathcal A[R_c + \rho] = \mathcal A_0(R_c)+ \frac{m}{R_c} \int_{0}^{2\pi} d\alpha \cdot \frac{1}{2}\big( \dot{\rho}^2 - 
 \rho^2\big)  + O(\rho^3)
= \frac{\pi}{\tilde\lambda} + \mathcal S[\rho;R_c] + O(\varrho^3)\,, \label{BubbleAction2}\\
\quad
\mathcal S[\rho;R_c] = \frac{m}{R_c} \int_{0}^{2\pi} \frac{1}{2}\big( \dot{\rho}^2 - \rho^2\big) d\alpha \,, \label{SBubbleAction}
\end{gather}
then \eqref{1BubbleCorrelator3} becomes
\[
G^{(1)}(L) = \frac{\pi L e^{-\tilde{M}_1 L}}{ \sqrt{L} \,\text{Vol}(\mathbb R^2)} e^{-\frac{\pi}{\tilde\lambda}}\,  R_c
\Big(I_1( N_1 R_c ) + L_{-1}( N_1 R_c ) - \frac{2}{\pi}\Big) \int \mathcal D \rho \, e^{-\mathcal S}  \,. \label{1BubbleCorrelator4}
\]
When the coefficient of quadratic term $\rho^2$ takes the value $\omega^2 = -1$, the action $\mathcal S$ of \eqref{SBubbleAction} has two extra zero modes, as:
\[
\rho (\alpha) \to \rho (\alpha) + a_x \cos \alpha + a_y \sin \alpha \,,
\]
which correspond to translations in two dimensions, and gives a divergence of $\mathcal D \rho$ proportional to $\text{Vol}(\mathbb R^2)$. This divergence cancels the divergent denominator in \eqref{1BubbleCorrelator4}.

With the quadratic action \eqref{SBubbleAction}, the functional integration $\mathcal D \rho$ reproduces a partition function of a periodic boson\footnote{Strictly speaking, with the negative coefficient of $\rho^2$ in \eqref{SBubbleAction}, the path integration \eqref{1BubbleCorrelator4} is a counting of metastable states with negative modes, and is defined only via analytical continuation \cite{Voloshin:1985id}. This fact introduces an extra factor of $i$ in the coming \eqref{DrhoRegularization}. Also, the extra factor of $\frac{1}{2}$ in \eqref{DrhoRegularization} is necessary, because there exist the negative mode $\rho = \text{Const}$, which corresponds to the variation of the bubble radius.}, and regularization of divergency gives\footnote{To regulate $\int \mathcal D \rho e^{-\mathcal S}$, add
$
\delta \mathcal S = \frac{\epsilon}{\pi} \int_{0}^{2\pi} d\alpha \rho^2 = 2\epsilon (a_x^2 +a_y^2)
$
to the action $\mathcal S$, with $\epsilon$ is the regulator. Denote $\omega$ as the "frequency" of the periodic boson, with $\delta \mathcal S$, the frequency shifts from $\omega^2 = -1$ to $\omega^2 = -1 + \frac{2 \epsilon R}{\pi m}$. The path integration now becomes convergent, with:
$$
\int \mathcal D \rho e^{-\mathcal S - \delta \mathcal S} = \frac{1}{2 \, | \sinh \pi \omega |} \sim \frac{m}{2 R}\frac{1}{\epsilon}\,.
$$
On the other hand, gaussian integration yields $\int da_x da_y e^{-2\epsilon (a_x^2 +a_y^2)} = \frac{\pi}{2 \epsilon} \to \text{Vol}(\mathbb R^2)$, and leads to \eqref{DrhoRegularization}.}:
\[
\int \mathcal D \rho \, e^{-\mathcal S[\rho ; R] }
= \frac{i}{2} \frac{m}{\pi R_c} \cdot \text{Vol}(\mathbb R^2) + (\text{finite terms}) \,. \label{DrhoRegularization}
\]
Combined with \eqref{DrhoRegularization}, \eqref{1BubbleCorrelator4} becomes:
\begin{gather}
G^{(1)}(L) = \frac{e^{-\tilde{M}_1 L}}{\sqrt{L}}\, m L  \cdot \frac{i}{2} \, \Big( I_1( N_1 R_c ) + L_{-1}( N_1 R_c ) - \frac{2}{\pi} \Big)
  e^{-\frac{\pi}{\tilde\lambda}} \,, \label{1BubbleCorrelator5}
\end{gather}
and from \eqref{Kappa_from_G1}, the instanton-like term reads:
\[
\mathcal K_1(\tilde\lambda) = - \frac{i}{2} \Big[ I_1( \frac{n_1(\tilde\lambda) }{ \tilde\lambda} ) + L_{-1}( \frac{n_1(\tilde\lambda) }{ \tilde\lambda}  ) - \frac{2}{\pi} \Big]
  e^{-\frac{\pi}{\tilde\lambda}} \,, \label{Kappa0full}
\]
where we have introduced $n_1(\tilde\lambda) = N_1/m$. The factor $e^{-\frac{\pi}{\tilde\lambda}}$ makes $\tilde\lambda = \frac{2\bar\sigma \mathfrak h}{m^2} \to 0^+$ an essential singularity of $\mathcal K_1(\tilde\lambda)$, and the prefactor is expandable in fractional powers of $\tilde\lambda$.

Denote the prefactor of \eqref{Kappa0full} as:
\[
 \mathcal I (z) = \Big( I_1( z ) + L_{-1}( z ) - \frac{2}{\pi} \Big) \,, \quad z = \frac{n_1(\tilde\lambda)}{\tilde\lambda} \,,
\]
by using the continuation of semiclassical spectrum (see Sec.\ref{Section2}), $ n_1(\tilde\lambda)$ can be expressed as:
\begin{gather}
n_1(\tilde\lambda) = 4\big[ \cosh\beta_1(\tilde\lambda) \cos \gamma_1(\tilde\lambda) - \cosh \vartheta_1(\tilde\lambda) \pm i \sinh\beta_1(\tilde\lambda) \sin \gamma_1(\tilde\lambda)  \big]\,, \label{n1def} \\
\text{with} \quad \sinh 2\vartheta_1 -  2\vartheta_1 = \frac{3}{2}\pi \tilde\lambda \,, \quad \sinh 2\tilde\vartheta_1 -  2\tilde\vartheta_1 = - \frac{3}{2}\pi \tilde\lambda \,,
\quad \tilde\vartheta_1 = \beta_1 \pm i \gamma_1 \,. \label{SCapproximation}
\end{gather}
Near $\tilde\lambda \to 0^+$, the function $n_1$ behaves as $n_1(\tilde\lambda) \sim \tilde\lambda^{2/3}$, thus $n_1 / \tilde\lambda$ diverges with:
\[
\frac{n_1(\tilde\lambda)}{\tilde\lambda} = -\sqrt{3} e^{\mp\frac{\pi i}{6}} \big( \frac{9\pi}{8} \big)^{\frac{2}{3}} \tilde\lambda^{-\frac{1}{3}} + O(\tilde\lambda^{\frac{1}{3}}) \,,
\quad \text{with} \quad a= \sqrt{3} \big( \frac{9\pi}{8} \big)^{\frac{2}{3}} \,. \label{n1overxidef}
\]
The expansion of $\mathcal I (z)$ at the large $z$ is: (denote $\omega = e^{\frac{\pi i}{3}}$)
\begin{gather}
\mathcal I (z) \xrightarrow{\tilde\lambda \to 0^+}
- \frac{2}{\pi} + \frac{\omega}{\pi a^2} \tilde\lambda^{\frac{2}{3}}
\Big[2 + 6 \big( \frac{\omega \tilde\lambda^{\frac{2}{3}}}{a^2} \big)
+ 90 \big( \frac{\omega \tilde\lambda^{\frac{2}{3}}}{a^2} \big)^2+ 3150 \big( \frac{\omega \tilde\lambda^{\frac{2}{3}}}{a^2} \big)^3
\cdots \Big]  \,, \quad \label{ILexpansion1}
\end{gather}
where $a$ was defined in \eqref{n1overxidef}
. Based on the expansion \eqref{ILexpansion1}, the leading contributions of $\mathcal K_1(\tilde\lambda)$ reads:
\[
\mathcal K_1(\tilde\lambda) = \frac{i}{\pi} e^{-\frac{\pi}{\tilde\lambda}} - \frac{16 i \, e^{\frac{\pi i}{3}} \tilde\lambda^{\frac{2}{3}}}{27 \cdot 3^{\frac{2}{3}}\pi^{\frac{7}{3}}}e^{-\frac{\pi}{\tilde\lambda}} + O(\tilde\lambda^{\frac{4}{3}}) \,, \label{Kappa0Leading}
\]
which is the instanton-like contribution to the mass of lightest particle in the metastable phase with $\xi \to 0^-$.

With similar strategies, one can compute the instanton-like contribution $\mathcal K_p(\tilde\lambda)$ for any heavier excitation with mass $M_p$, by modifying the quantization condition \eqref{SCapproximation} as $\frac{3}{2} \pi \tilde\lambda \to 2 \pi \tilde\lambda (p - \frac{1}{4})$. The first term of $\mathcal K_p(\tilde\lambda)$ expansion is $\frac{i}{\pi} e^{-\frac{\pi}{\tilde\lambda}} $, which is identical to the one of \eqref{Kappa0Leading}, because it comes from area of the cigar $\mathcal C(L,R_c)$. Subleading terms of $\mathcal K_p(\tilde\lambda)$ have similar patterns comparing to \eqref{Kappa0Leading}, with different coefficients.

Finally, we back to the instanton-like contribution to the analysis of Sec.\ref{Section2} and Sec.\ref{Section3}. Denote the universal leading contribution of $\mathcal K_p(\tilde\lambda)$ as ($x = -\xi > 0$):
\[
\mathcal K_0(\tilde\lambda) = \frac{i}{\pi} e^{-\frac{\pi}{\tilde\lambda}}  = \frac{i}{\pi} e^{ -\frac{\pi}{2 \bar s x} } \,,  \label{Kappa0Expression}
\]
which should contribute to the discontinuities \eqref{FLdisc} at small negative $\xi$ along the upper/lower edges of the Fisher-Langer's branch cut. The leading correction to $\Im m \, \hat M_p( \xi + i 0)$ is:
\[
\Im m \, \mathcal K_0(\tilde\lambda) = \frac{1}{\pi} e^{-\frac{\pi}{\tilde\lambda}} = \frac{1}{\pi} e^{ -\frac{\pi}{2 \bar s x} }  \, , \label{ImKappa0Expression}
\]
which is as shown as the near horizontal magenta line in Fig.\ref{M123_LowT_discontinuities_piecewise1}. Obviously, comparing to the discontinuities from continuation (\eqref{Im_Mp_expansion_mn} and \eqref{semixdisc}), the contribution \eqref{ImKappa0Expression} is negligible.

On the complex $\eta$ plane, the instanton-like contribution \eqref{Kappa0Leading} should follow the continuation across the shadow domain
. At the edges of low-T wedge $\text{Arg}\, \eta = \pm \frac{8}{15}\pi$, at large $|\eta|$ the $\mathcal K_1$ term contributes as:
\[
\mathcal M_1(\eta) \to \mathcal M_1^{\text{exp}}(\eta = z e^{\pm \frac{8\pi i}{15}}) + \hat{\mathcal K}_1(z) \,, \quad
\hat{\mathcal K}_1(z) \to \frac{i}{\pi} e^{\pm\frac{8}{15}\pi} z e^{ -\frac{\pi}{2 \bar s} z^{\frac{15}{8}}}\,, \label{HatKappaDef}
\]
where $z = e^{\mp \frac{8\pi i}{15}}\eta$, and $\mathcal M_1^{\text{exp}} $ is the expansion of \eqref{M1etaExpansionLowT1} with $\eta = z e^{\pm \frac{8\pi i}{15}}$. As illustrated in Fig.\ref{eta_planes}, the continuation across the shadow domain is achieved by rotating $\eta$ for $\frac{\pi}{5}$, as from the Fisher-Langer's branch cuts $\text{Arg}\, \eta = \pm \frac{8}{15}\pi$ to the rays of Yang-Lee branch cut $\text{Arg}\, \eta = \pm \frac{11}{15}\pi$.

Qualitatively, to obtain an exponential growth the exponent of \eqref{Kappa0Leading} must have a positive real part. In other words, \eqref{Kappa0Leading} changes its behaviours to oscillating growth after rotating $\xi$ further for $\frac{\pi}{2}$ inside the Fisher-Langer's branch cut, or equivalently rotating $\eta$ for $\frac{4}{15}\pi$ towards the shadow domain. With the angular size of shadow domain is $\Omega^{(\eta)}_{\text{SD}} =  \frac{\pi}{5}$, the Stokes phenomenon of $\mathcal K_1$ term is hiding inside the Yang-Lee branch cut of Fig.\ref{eta_planes}(b), and the $M_1$ extended analyticity is safe from extra singularities along the Yang-Lee branch cut.

Though by rotating $z \to y e^{\frac{\pi}{5}}$, \eqref{HatKappaDef} is exponentially small at $y \to +\infty$, quantitatively it's necessary to show that the contribution of \eqref{HatKappaDef} at finite $y$ is sufficiently small, and the approximated $\Delta_1(y)$ from Appendix \ref{Appendix_B} remains intact. After rotating $z \to y e^{\frac{\pi}{5}}$, the contribution of $\Re e\, \hat{\mathcal K}_1(y)$ and $\Im m\, \hat{\mathcal K}_1(y)$ are as shown in Fig.\ref{Continued_Disc_Delta1}, with two solid lines which almost indistinguishable from the horizontal axis. As a result, the instanton-like terms of \eqref{HatKappaDef} are negligible on the interval $Y_0<y<+\infty$, and can be ignored in the interpolation approach discuss in Appendix \ref{Appendix_B}.

Finally, despite that the instanton-like terms \eqref{Kappa0Leading} and \eqref{HatKappaDef} only have tiny contribution to the low-T analyticity and extended analyticity of $M_1$, they are implying something hiding within the Yang-Lee branch cut of Fig.\ref{eta_planes}(b). The rotation of $\eta$ for $\frac{4}{15}\pi$ towards the shadow domain would lead to the oscillating growth of $\hat{\mathcal K}_1$, and it suggests that the ray $\text{Arg}\, \eta = \pm \frac{4}{5}\pi$ may have interesting analyticity structures. One of the conjecture is that there exist an infinite number of singularities on the second sheet of $\eta$, within the Yang-Lee branch cut of Fig.\ref{eta_planes}(b), and these singularities are condensing towards $|\eta| \to \infty$ along the ray $\text{Arg}\, \eta = \pm \frac{4}{5}\pi$. We will provide further detailed analysis in future works.

\bibliography{isingref}

\begin{thebibliography}{10}

\bibitem{Xu:2022mmw}
Hao-Lan Xu and Alexander Zamolodchikov.
\newblock {2D Ising Field Theory in a magnetic field: the Yang-Lee
  singularity}.
\newblock {\em JHEP}, 08:057, 2022.

\bibitem{fonseca2003ising}
P~Fonseca and A~Zamolodchikov.
\newblock Ising field theory in a magnetic field: analytic properties of the
  free energy.
\newblock {\em Journal of statistical physics}, 110(3-6):527--590, 2003.

\bibitem{zamolodchikov2013ising}
Alexander Zamolodchikov.
\newblock Ising spectroscopy ii: particles and poles at t> tc.
\newblock {\em arXiv preprint arXiv:1310.4821}, 2013.

\bibitem{Xu:2023nke}
Hao-Lan Xu and Alexander Zamolodchikov.
\newblock {Ising Field Theory in a magnetic field: $\varphi^3$ coupling at $T >
  T_c$}.
\newblock 4 2023.

\bibitem{BazhanovYL}
Vladimir~V. Mangazeev, Bryte Hagan, and Vladimir~V. Bazhanov.
\newblock {Corner Transfer Matrix Approach to the Yang-Lee Singularity in the
  2D Ising Model in a magnetic field}.
\newblock 8 2023.

\bibitem{yang1952statistical}
Chen-Ning Yang and Tsung-Dao Lee.
\newblock Statistical theory of equations of state and phase transitions. i.
  theory of condensation.
\newblock {\em Physical Review}, 87(3):404, 1952.

\bibitem{lee1952statistical}
Tsung-Dao Lee and Chen-Ning Yang.
\newblock Statistical theory of equations of state and phase transitions. ii.
  lattice gas and ising model.
\newblock {\em Physical Review}, 87(3):410, 1952.

\bibitem{cardy1985conformal}
John~L Cardy.
\newblock Conformal invariance and the yang-lee edge singularity in two
  dimensions.
\newblock {\em Physical review letters}, 54(13):1354, 1985.

\bibitem{mccoy2013two}
Barry~M McCoy and Tai~Tsun Wu.
\newblock {\em The two-dimensional Ising model}.
\newblock Harvard University Press, 2013.

\bibitem{fisher1978yang}
Michael~E Fisher.
\newblock Yang-lee edge singularity and $\phi$ 3 field theory.
\newblock {\em Physical Review Letters}, 40(25):1610, 1978.

\bibitem{Zamolodchikov:1990bk}
A.B. Zamolodchikov.
\newblock {Two point correlation function in scaling Lee-Yang model}.
\newblock {\em Nucl. Phys. B}, 348:619--641, 1991.

\bibitem{Cardy:1989fw}
John~L. Cardy and G.~Mussardo.
\newblock {S Matrix of the Yang-Lee Edge Singularity in Two-Dimensions}.
\newblock {\em Phys. Lett.}, B225:275--278, 1989.

\bibitem{smirnov2017space}
FA~Smirnov and AB~Zamolodchikov.
\newblock On space of integrable quantum field theories.
\newblock {\em Nuclear Physics B}, 915:363--383, 2017.

\bibitem{Zamolodchikov:1995xk}
Alexei~B. Zamolodchikov.
\newblock {Mass scale in the sine-Gordon model and its reductions}.
\newblock {\em Int. J. Mod. Phys. A}, 10:1125--1150, 1995.

\bibitem{Zamolodchikov:1989fp}
A.B. Zamolodchikov.
\newblock {Integrals of Motion and S Matrix of the (Scaled) T=T(c) Ising Model
  with Magnetic Field}.
\newblock {\em Int. J. Mod. Phys. A}, 4:4235, 1989.

\bibitem{Fateev:1993av}
V.~A. Fateev.
\newblock {The Exact relations between the coupling constants and the masses of
  particles for the integrable perturbed conformal field theories}.
\newblock {\em Phys. Lett. B}, 324:45--51, 1994.

\bibitem{Delfino:1996xp}
G.~Delfino, G.~Mussardo, and P.~Simonetti.
\newblock {Nonintegrable quantum field theories as perturbations of certain
  integrable models}.
\newblock {\em Nucl. Phys. B}, 473:469--508, 1996.

\bibitem{Fateev:1997yg}
Vladimir Fateev, Sergei~L. Lukyanov, Alexander~B. Zamolodchikov, and Alexei~B.
  Zamolodchikov.
\newblock {Expectation values of local fields in Bullough-Dodd model and
  integrable perturbed conformal field theories}.
\newblock {\em Nucl. Phys. B}, 516:652--674, 1998.

\bibitem{Alekseev:2011my}
Oleg Alekseev.
\newblock {Form factors in the Bullough-Dodd related models: The Ising model in
  a magnetic field}.
\newblock {\em JETP Lett.}, 95:201--205, 2012.

\bibitem{fisher1967theory}
Michael~E Fisher.
\newblock The theory of condensation and the critical point.
\newblock {\em Physics Physique Fizika}, 3(5):255, 1967.

\bibitem{langer2000theory}
James~S Langer.
\newblock Theory of the condensation point.
\newblock {\em Annals of Physics}, 281(1-2):941--990, 2000.

\bibitem{andreev1964singularity}
AF~Andreev.
\newblock Singularity of thermodynamic quantities at a first order phase
  transition point.
\newblock {\em Sov. Phys. JETP}, 18:1415--1416, 1964.

\bibitem{gunther1980goldstone}
NJ~Gunther, DJ~Wallace, and DA~Nicole.
\newblock Goldstone modes in vacuum decay and first-order phase transitions.
\newblock {\em Journal of Physics A: Mathematical and General}, 13(5):1755,
  1980.

\bibitem{lowe1980instantons}
MJ~Lowe and DJ~Wallace.
\newblock Instantons and the ising model below tc.
\newblock {\em Journal of Physics A: Mathematical and General}, 13(10):L381,
  1980.

\bibitem{harris1984ising}
CK~Harris.
\newblock The ising model below tc: calculation of nonuniversal amplitudes
  using a primitive droplet model.
\newblock {\em Journal of Physics A: Mathematical and General}, 17(3):L143,
  1984.

\bibitem{Voloshin:1985id}
M.~B. Voloshin.
\newblock {DECAY OF FALSE VACUUM IN (1+1)-DIMENSIONS}.
\newblock {\em Yad. Fiz.}, 42:1017--1026, 1985.

\bibitem{Wu:1975mw}
Tai~Tsun Wu, Barry~M. McCoy, Craig~A. Tracy, and Eytan Barouch.
\newblock {Spin spin correlation functions for the two-dimensional Ising model:
  Exact theory in the scaling region}.
\newblock {\em Phys. Rev. B}, 13:316--374, 1976.

\bibitem{mccoy1978two}
Barry~M McCoy and Tai~Tsun Wu.
\newblock Two-dimensional ising field theory in a magnetic field: Breakup of
  the cut in the two-point function.
\newblock {\em Physical Review D}, 18(4):1259, 1978.

\bibitem{tHooft:1974pnl}
Gerard 't~Hooft.
\newblock {A Two-Dimensional Model for Mesons}.
\newblock {\em Nucl. Phys. B}, 75:461--470, 1974.

\bibitem{Fonseca:2006au}
Pedro Fonseca and Alexander Zamolodchikov.
\newblock {Ising spectroscopy. I. Mesons at T \ensuremath{<} T(c)}.
\newblock 12 2006.

\bibitem{Rutkevich:2009zz}
S.~B. Rutkevich.
\newblock {Formfactor perturbation expansions and confinement in the Ising
  field theory}.
\newblock {\em J. Phys. A}, 42:304025, 2009.

\bibitem{fonseca2006ising}
Pedro Fonseca and Alexander Zamolodchikov.
\newblock Ising spectroscopy i: Mesons at t< t\_c.
\newblock {\em arXiv preprint hep-th/0612304}, 2006.

\end{thebibliography}

\end{document}